\documentclass[journal,comsoc]{IEEEtran}
\pdfoutput=1

\usepackage[T1]{fontenc}

\ifCLASSINFOpdf
  \usepackage[pdftex]{graphicx}
  \graphicspath{{./figures/}}
  \DeclareGraphicsExtensions{.pdf,.jpg,.png,.xps}
\else
  \usepackage[dvips]{graphicx}
  \graphicspath{{./}}
  \DeclareGraphicsExtensions{.eps}
\fi

\usepackage{cite}
\usepackage{amsmath} 
\usepackage[cmintegrals]{newtxmath}
\usepackage{booktabs} 
\usepackage{url} 
\usepackage{soul, color} 
\usepackage{algorithm}
\usepackage[noend]{algpseudocode}
\usepackage{float}
\usepackage{multirow}
\ifCLASSOPTIONcompsoc
 \usepackage[caption=false,font=normalsize,labelfont=sf,textfont=sf]{subfig}
\else
 \usepackage[caption=false,font=footnotesize]{subfig}
\fi

\begin{document}

\title{On Optimal Placement of Hybrid Service Function Chains (SFCs) of Virtual Machines and Containers in a Generic Edge-Cloud Continuum}

\author{Francisco~Carpio,~Wolfgang~Bziuk~and~Admela~Jukan
	\thanks{F. Carpio, W.Bziuk and A.Jukan are with the Institute of Computer and Network Engineering, Technische Universit{\"a}t Braunschweig, Germany. E-mail: \{f.carpio, w.bziuk, a.jukan\}@tu-bs.de.}
}


\maketitle

\newcommand{\vmvnf}{VM-VNF}
\newcommand{\ctvnf}{CT-VNF}
\newcommand{\vmonly}{vm-only}
\newcommand{\ctonly}{ct-only}
\newcommand{\hybrid}{vm-ct}

\begin{abstract}
	Traditionally, Network Function Virtualization (NFV) has been implemented to
	run on Virtual Machines (VMs) in form of Virtual Network Functions (VNFs).
	More recently, the so-called Serverless Computing has gained traction in
	cloud computing, offering Function-as-a-Service (FaaS) platforms that make
	use of containerization techniques to deploy services. In contrast to
	VM-based VNFs, where resources are usually reserved and continuously
	running, FaaS can just be subsets of code implementing small functions
	allowing for event-driven, on-demand instantiations. Thus, a hybrid
	VM-Container based Service Function Chains (SFCs) are a natural evolution of
	NFV architecture. We study a novel problem of optimal placement of hybrid
	SFCs from an Internet Service Provider (ISP) point of view, whereby VNFs can
	be instantiated either over VMs or containers in a generic edge and cloud
	continuum. To this end, we propose a Mixed-Integer Linear Programming model
	as well as a heuristic solution to solve this optimization problem that
	considers three objectives unique to the specific VM and container
	deployment in a carrier network: operational costs for maintaining servers
	in the edge, costs of placing VNFs in third-party cloud providers and
	penalty costs applied when SLA agreements are violated in terms of
	end-to-end delay. We also propose 2-phases optimization process to analyze
	the effect on performance as a result of replications and migrations of
	VNFs. The model can be used to highlight scenarios where a combination of
	VMs and containers can provide most benefits from the monetary costs point
	of view.
\end{abstract}


\IEEEpeerreviewmaketitle

\section{Introduction}

\IEEEPARstart{I}{nternet} Service Providers (ISP) recognize Network Function
Virtualization (NFV) as a key concept to reducing capital and operational
expenditures (Capex and Opex). In NFV, Virtual Network Functions (VNFs) run over
Virtual Machines (VMs) that are managed in the Network Function Virtualization
Infrastructure (NFVI) by the NFV Management and Orchestrator module (NFV
MANO)\cite{ETSI2016}. Full service provisioning is achieved by concatenating
multiple VNFs in an specific sequence order, defined as Service Function Chains
(SFCs). The placement of chained VNFs into physical servers, known as \emph{VNF
	placement problem,} can follow different optimization objectives, such as
network load balancing, reliability, end-to-end delay, etc. More recently, NFV
concept has further evolved towards the so-called \emph{Serverless Computing},
adopting Function-as-a-Service (FaaS) technology as a way to also deploy VNFs
over containers, with AWS Lambda being the first public cloud Infrastructure to
offer these services in 2014 \cite{aws}. To further reduce Opex, it is expected
that traditional VM-based VNFs (\vmvnf s) are going to co-exist with
container-based VNFs (\ctvnf s) under the same management system. The Open
Source Mano (OSM) is already supporting the integration of Virtual
Infrastructure Managers (VIMs), a serverless platform that can integrate VNFs,
such as OpenWhisk \cite{Alvarez2019b}.

So far, little is known about placement of VNFs that run over containers (\ctvnf
s), and especially when chained together with VMs in an SFC. From the point of
view of the server utilization, \vmvnf s require much more server resources
compared to \ctvnf s, due to the overhead introduced by the hypervisor-based
virtualization technology. From the OPEX point of view, \ctvnf s deployed on
third party cloud services are more expensive as they are usually charged by
usage, as opposed to VMs that typically have fixed cost rates by deployment.
Furthermore, NFV can make use of migrations and replications of VNFs, for
instance, to adapt to service request variability, to increase quality of
service (QoS), reliability, etc. While migrations are known to increase the
probability of violating SLAs due to service interruptions incurring into
penalty costs for the ISP, replications require extra server and network
resources due to the overhead and state synchronization. If an ISP needs to rent
resources from a cloud provider, on the other hand, the choice between \vmvnf s
and \ctvnf s has direct impact on the costs to pay due to the different charging
rates of a cloud provider. Thus, the optimization of \vmvnf s and \ctvnf s
placement presents a few important and interesting challenges which have not
been solved yet. This is especially interesting to solve in an ISP-centric model
of edge-to-cloud continuum, where it can be differentiated between the costs
associated with the edge network and the cloud considering that the ISP owns the
edge infrastructure and is charged when using a third-party cloud.

In this paper, we propose a Mixed-Integer Linear Programming (MILP) model to
study the placement optimizations of VMs and containers in hybrid SFC
configurations in a generic ISP-centric model of edge-cloud continuum. We
propose to solve the VNF placement problem in two phases, starting off with low
traffic load and solving the placement problem under high traffic load. We use
the low traffic phase as an initial placement to understand the effects in the
second phase of, from one side, migrations (that have on the penalty costs
caused by service interruptions) and, on the other side, replications (that
impact the network and server resources due to state synchronization tasks). To
this end, we define a joint optimization problem with three different costs
models for the ISP: (i) operational costs for maintaining servers in the edge
network, (ii) charges applied when the service provider uses a third party cloud
provider, and (iii) penalty costs applied for SLA violations. In contrast to
other placement models based on variants of well-known multi-commodity flow
problem, our model optimally uses multiple traffic flows in SFCs, and unlike any
known SFC multipath model that only restrict the propagation delay of all
predefined paths or use the end-to-end delay simply averaged over paths, our
model considers individual end-to-end delays for each traffic flow. We also
propose a greedy algorithm as an online solution that performs close to the
optimal solution. The results show the performance and the related
cost-tradeoffs of hybrid SFC deployment.

The rest of the paper is organized as follows. Section II presents related work.
Section III describes the system model. Section IV formulates the optimization
model. Section V describes the heuristics. Section VI analyzes the performance
and Section VII concludes the paper.

\section{Related Work and Our Contribution}

\subsection{VNF placement, migrations and replications}

Significant amount of previous work has focused on the VNF placement problem
considering VMs as underlying technology. More in detail, \vmvnf s placement
problem has been addressed by the research community with variants of the joint
optimization placement problem with different objectives. For instance, in
\cite{Tajiki2017}, a resource allocation solution is proposed for optimizing
energy efficiency, while considering delay, network and server utilization.
\cite{Wang2016a} proposes a joint optimization solution, considering network and
service performance, to not only solve the VNF placement problem, also known as
Forwarding Graph Embedding, but also the VNF Chain Composition and VNFs
scheduling problems. \cite{Basta2017} proposed models to finding the optimal
dimensioning and resource allocation with latency constraints in mobile
networks.

Whereas no distinctive challenges have been reported between VMs and container
placement \cite{Laghrissi2019}, considering migrations and replication, the two
concepts exhibit more features that need to be differentiated. In case of \vmvnf
s, and under certain circumstances, migration is required, which can be
implemented either by migrating the entire VM \cite{Xia2016} or by migrating
only the internal states of an VNF \cite{Xia2016a}. Accordingly, multiple
objectives can be applied, such as minimizing the impact on QoS, maximizing the
energy efficiency, maximizing load balancing, etc (e.g.
\cite{Gember-Jacobson2014}, \cite{Ghaznavi2015} and \cite{Hawilo2017}). In this
regard, it has been reported that the migration process can be up to four times
faster with containers than when using VMs \cite{Zou2018}. Since the main issue
with \vmvnf s migrations is the problem of service interruptions that negatively
impact QoS, previous work, e.g., \cite{Eramo2017} derived a trade-off between
the power consumption and QoS degradation to determine whether a migration is
appropriate. On the other hand, \vmvnf s replications (sometimes also referred
as \emph{backups}) have been primarily used to provide service reliability
\cite{michael2016} while trying to minimize the number of required replicas
\cite{Ding2017}, to reduce end-to-end service delays \cite{Qu2017}
\cite{Alleg2018a}, to load balance the link utilization \cite{Carpio2017a} and
to load balance the server utilization \cite{Carpio2017b}. While in a related
work in \cite{Huang2018}, the authors try to balance the number of migrations
and replications in order to maximize the network throughput and minimize the
end-to-end delay, in \cite{Carpio2018} the objective is to reduce migrations by
using replications to find a trade-off between both mechanisms to improve
server, network load balancing and QoS. In this regard, when using containers to
create VNF replicas, it is proven that they can reduce instantiation delay and
to increase throughput while achieving near zero downtime of the SFC
\cite{Filipe2019}.

\subsection{Edge-Cloud Continuum and the Issues of Placement}

The emerging edge computing trend is looking at distributing computation
resources closer to the network users in order to minimize end-to-end latencies
and to reduce the transit traffic in the core network. In these highly
distributed edge scenarios, where the computing capabilities are typically more
constrained than in the cloud, the standard full MANO stack is not suitable.
Instead, new solutions propose the usage of lightweight virtualization platforms
using containerization techniques as a solution for orchestration and management
of network services \cite{Riggio2018}, \cite{Sewak2018} and for the deployment
of VNFs achieving very low resource usage overhead, almost comparable to bare
metal \cite{Sheoran2017}. The latest developments in the Open Source Mano
already include support not only for virtual, physical and hybrid network
functions (VNFs, PNFs and HNFs, respectively), but also to Virtual
Infrastructure Managers (VIMs) and WAN Infrastructure Managers (WIMs)
\cite{OSM2019}. One of the recent projects making use of the VIM support
integrates Apache OpenWhisk which has support to FaaS, through the development
of a VIM plugin \cite{Alvarez2019b}. In this context, more recent studies, e.g.,
\cite{Cziva2018}, analyze the VNF placement at the edge in order to improve QoS
by constraining the maximum number of SLA violations in terms of end-to-end
latency, which is the salient feature of edge computing. In \cite{Soni2017}, a
management mechanism is proposed, based on NFV/SDN ISP networks to provide
multicast edge-based services. Similarly, in \cite{Boubendir2017}, a deployment
model of VNFs is presented with location-awareness to meet the QoS in mobile
networks.

\subsection{Our Contribution}

Motivated by the fact that NFV-enabled edge-computing networks need to be
optimized in terms of profit, as shown in \cite{Ma2018} or \cite{Racheg2017},
and inspired by related work that not only optimize operational costs but also
resource utilization, while making sure that service level agreements are met,
as in \cite{Bari2016}, we contribute with a model that minimizes \emph{costs} as
an objective function when placing VNFs in a generic edge-cloud continuum. In
our model, we consider the specific constraints arising from a hybrid SFCs of
VMs and containers, which has not been addressed yet. We analyze this problem
from an ISP point of view by using a MILP multipath based model that minimizes
three different cost functions when placing VNFs: (i) costs for maintaining the
servers deployed at the edge network, (ii) cost applied when using third-party
cloud infrastructure to deploy VNFs, and (iii) penalty costs applied in case of
SLA violations. This model also considers the impact of migration of VNFs on the
penalty costs due to service interruptions as well as the impact of replications
of VNFs on the OPEX due to the extra server and network resources required. For
the calculation of the penalty costs applied when maximum service delay is
exceed, we consider path delay constraints on a per-path basis, i.e.,
individually per each path, and we also consider end-to-end delays of each
traffic demand where service interruptions delays are caused by migrations.
Unlike previous work, we consider the impact of synchronization traffic in the
network used for maintaining state synchronization between VNF replicas. Since
the proposed model does not scale due to complexity when used on a comparatively
large network, we propose a heuristic approach as an online solution.

\begin{figure}[!t]
	\centering
	\includegraphics[width=1.0\columnwidth]{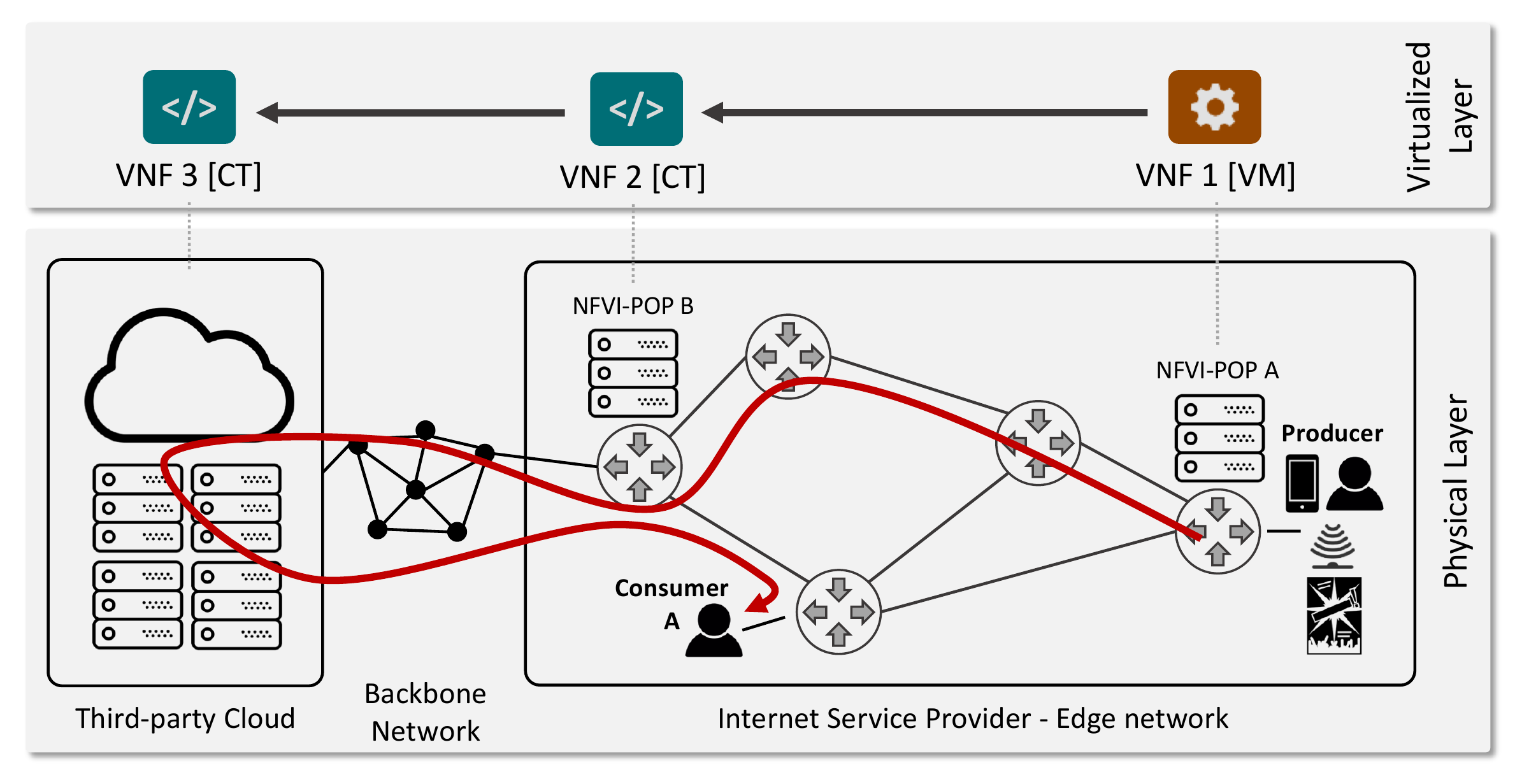}
	\caption{Example of hybrid SFC in an edge to cloud scenario}
	\label{system}
\end{figure}

\begin{figure*}[!ht]
	\centering
	\subfloat[Initial placement]{\includegraphics[width=0.33\textwidth]{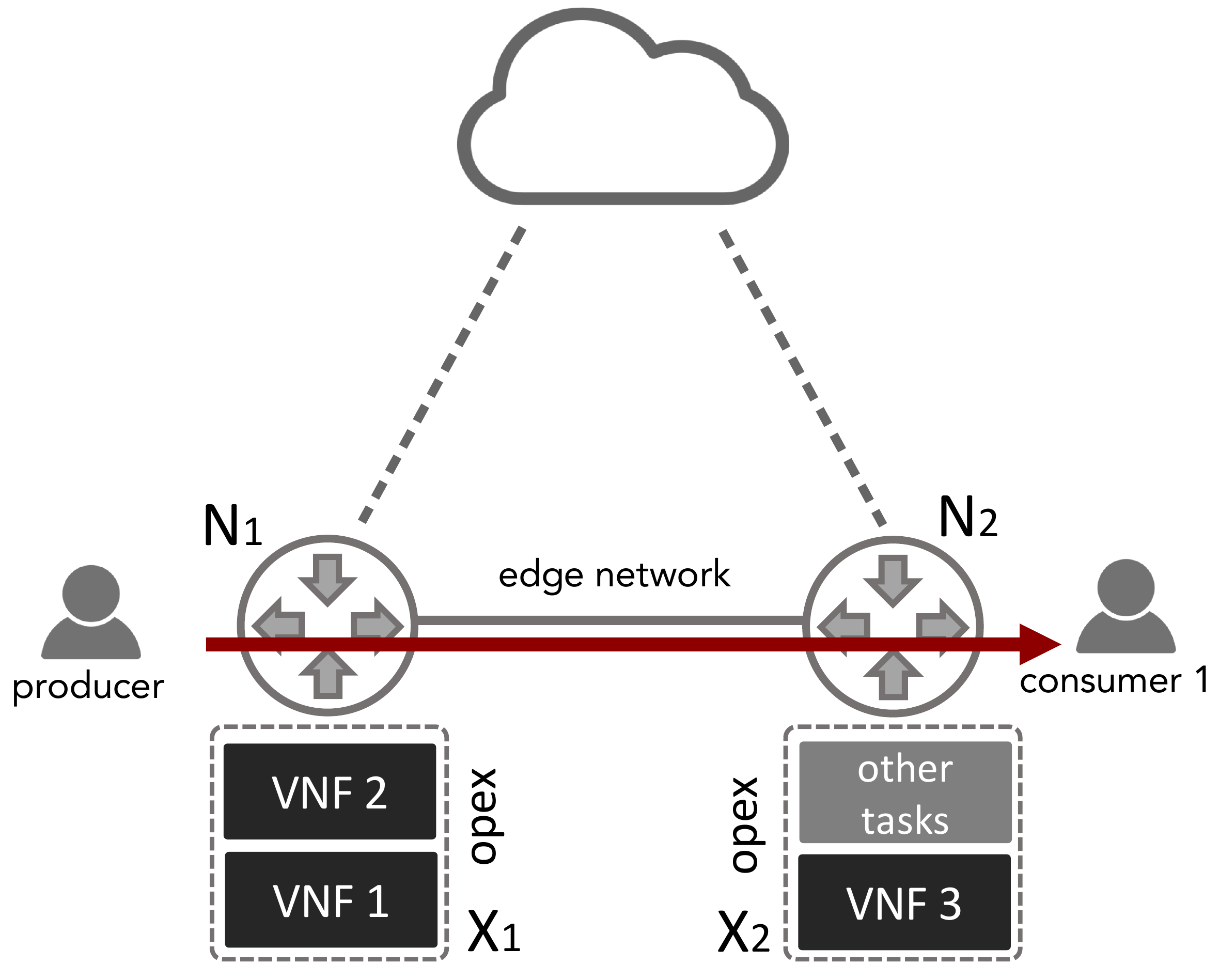}%
		\label{model_init}}
	\hfil
	\subfloat[Migration]{\includegraphics[width=0.33\textwidth]{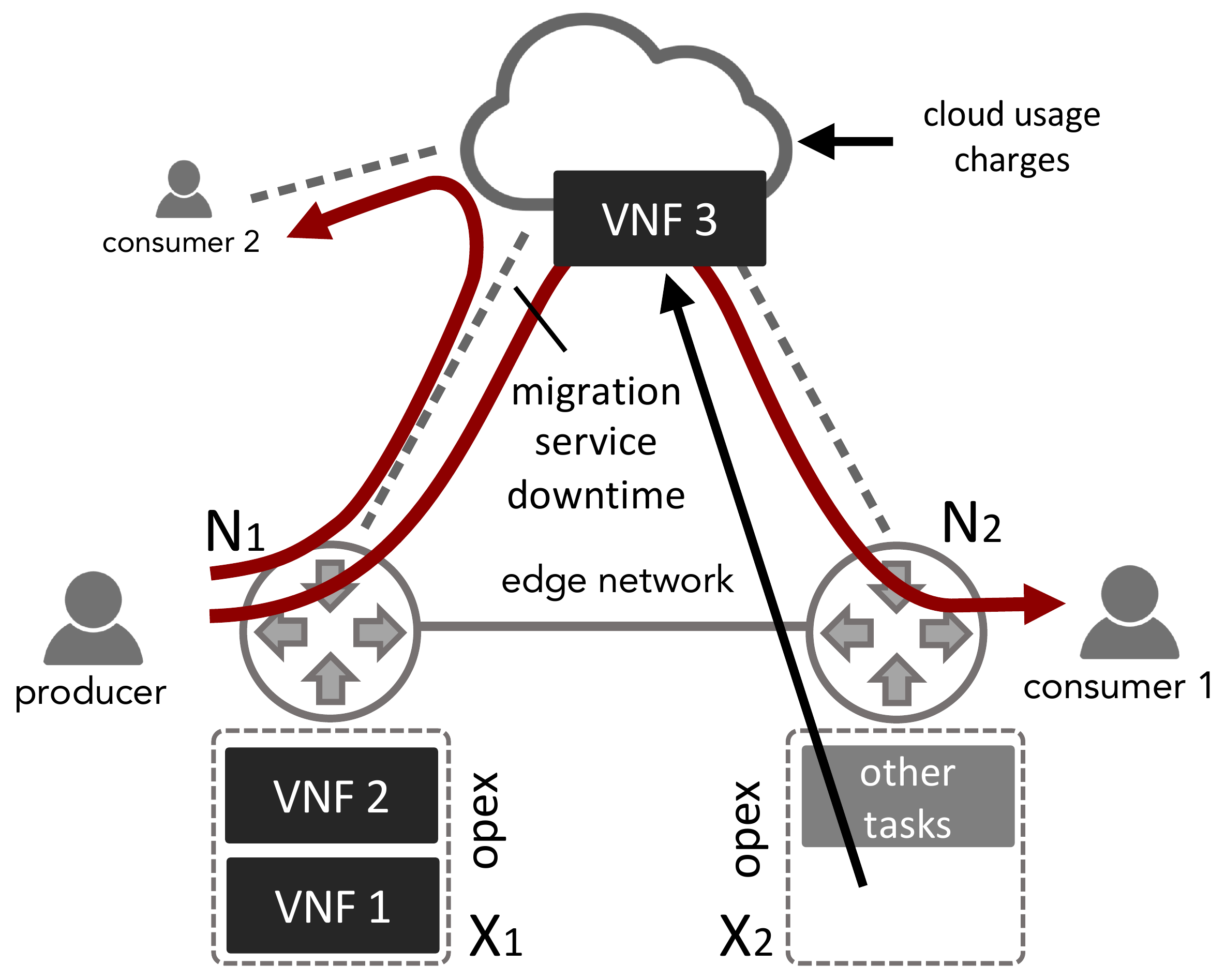}%
		\label{model_mgr}}
	\hfil
	\subfloat[Replication]{\includegraphics[width=0.33\textwidth]{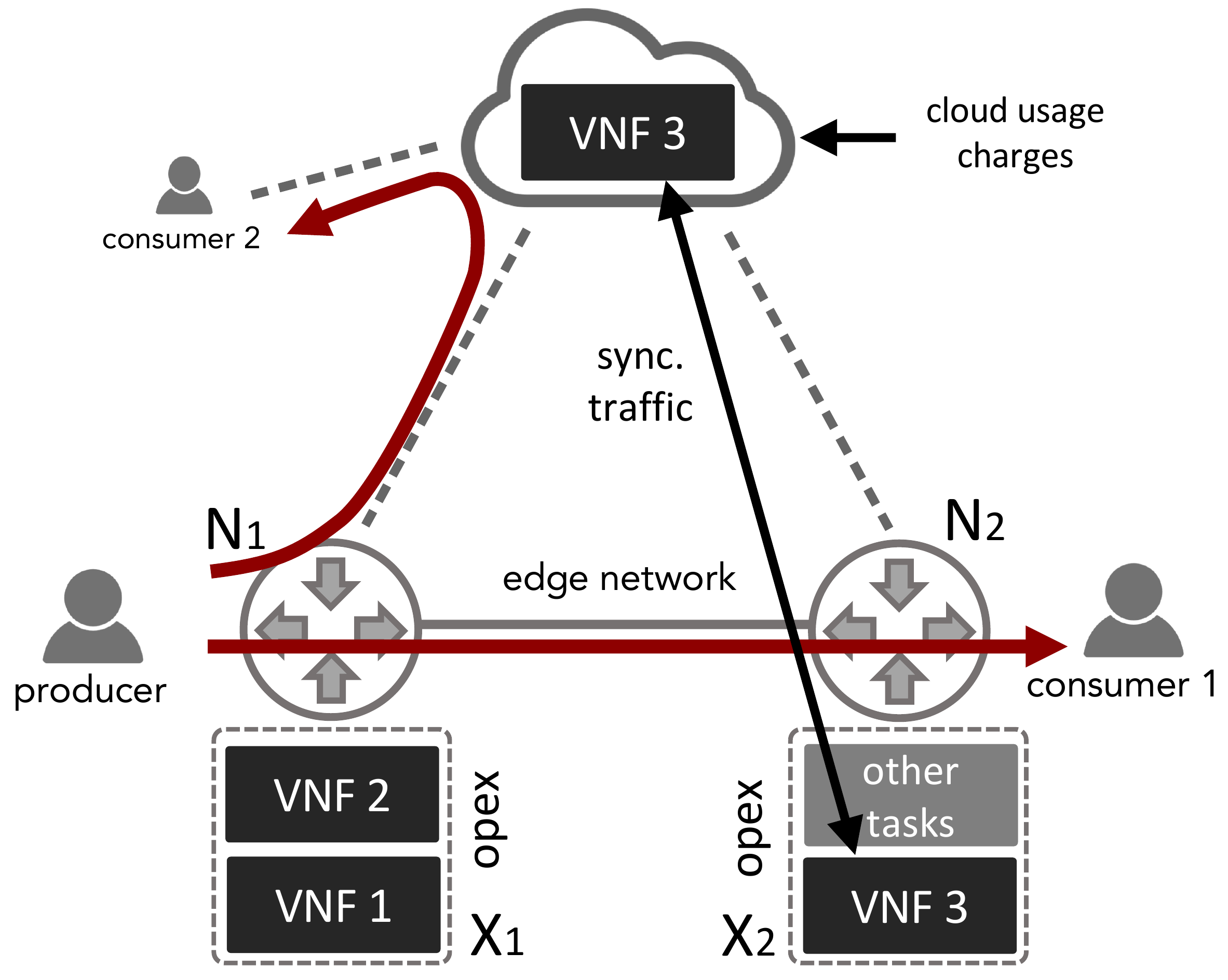}%
		\label{model_rep}}
	\caption{Migration and replication examples after an initial placement}
	\label{models}
\end{figure*}

\section{System Model} \label{system_model}

\subsection{Reference Scenario}

Fig. \ref{system} illustrates the reference scenario. We assume that an ISP owns
the network infrastructure close to the end users where it deploys small groups
of servers for the NFV Infrastructure as Point-of-Presence (NFVI-POP) to run
VNFs in an edge computing scenario. We also assume that the ISP uses the cloud
as a third party to offload the VNFs when necessary. Let us now illustrate a
hybrid SFC of three VNFs placed at different points in the network and cloud.
Here, the ISP has to optimize the placement of VNFs while minimizing the
operational costs of their deployment. In such scenarios, containerization
technologies enhance NFV with a faster and more flexible way of provisioning the
new services. Service providers can also flexibly place VNFs in either their own
premises (i.e., at the network edge) or in third party cloud provider
infrastructure. The decision on where to place VNFs is economy-driven. For
instance, whereas placing VNFs in-network edge premises incurs energy and
hardware maintenance costs, placing them on third party cloud providers adds
additional charges that vary depending if they are deployed as VMs, usually
charged per amount of used resources, or as containers, usually charged by
duration of time. In both cases, the SLA violations need to be considered as
penalty costs, which every ISP and cloud provider try to minimize.

\subsection{VNF Migrations and Replications}

To illustrate the need for migrations or replications (be it for \vmvnf s  and
\ctvnf s), let us consider an example where three VNFs are chained and providing
service to the traffic generated from one producer to one consumer. Let us
consider a service provider edge network with two nodes, N1 and N2, with one
server each $X_1$ and $X_2$, see Fig.\ref{model_init}. Server $X_1$ allocates
VNFs 1 and 2, using all the resources in the server, while server $X_2$
allocates VNF 3, leaving the spare capacity to be used by other tasks. Now,
let's assume there is another consumer located in a different network (see Fig.
\ref{model_mgr}). In this case, since there are no free resources to serve the
new service request, one option is to migrate VNF 3 to the cloud. This process
is usually carried out by instantiating the new VNF and synchronizing the state
before shutting shutting down the old one. However, there is a short service
downtime caused by the control mechanisms for rerouting the traffic of the
affected flows to the new location. This additional delay caused by the
migration can affect the QoS and cause penalty costs to the service provider
when the maximum delay specified in the SLA agreement is exceeded. On the other
hand, there is also the possibility of replicating VNF 3 into the cloud without
redirecting the original traffic, as shown in Fig. \ref{model_rep}. In this
case, only the new traffic to the second consumer is redirected to the cloud to
be processed by VNF 3. In this case, since we assume stateful VNFs, we need to
consider the synchronization traffic between the original VNF and the replica.
This example illustrates three basic costs to be considered when placing VNFs in
the network or in the cloud: operational costs for maintaining servers in the
edge network, additional charges when placing VNFs in the cloud and penalty
costs when the service delay specified in the SLA agreement violated.

\subsection{Assumptions}

For the sake of modeling and independently of placing VNFs in the edge or in the
cloud, there are some assumptions to be considered here. First, when deploying
VNFs over VMs in a model, one VNF instance maps 1:1 to a VM where some server
resources are reserved to the VM to run independently of the processed traffic.
This overhead is not considered when deploying VNFs as containers, since their
performance is very close to bare metal. Regarding to delay, we define the
end-to-end service delay, as the sum of propagation delay (time for the data to
travel trough the fiber), processing delay (time for the VNF to process the
data) and service interruption delays caused by migrations. Since before
performing a migration, the new VNF in the new location has to be instantiated
to then synchronize the states with the old VNF, the delay that affects to the
service quality is the one related only to the short interruption of the active
flows to commute and is independent from the VM or container size
\cite{Taleb2019}. Although this delay is just temporal, we consider it as a key
aspect to determine at some point in time whether a VM migration exceeds the
maximum allowed service delay in the SLA agreement or not. In this sense, the
service delay can be interpreted as a worst case delay. Since our model is
multipath based, every service chain can use multiple paths, whereby each path
can exhibit different delays due to different links traversed and the related
VNFs. Regarding to the use of replications, it should be noted that in our model
replications are used to address scalability during high load periods, and not
for reliability purposes. We also consider the additional traffic generated by
replicas to maintain their state synchronization independently if they run over
VMs or containers. Unlike migrations, replicas do not induce extra service
delay, since ongoing traffic demands are never interrupted during the
replication process.

\subsection{Optimization Scenarios} \label{opt_scenarios}

Our approach to optimizations is carried out from the point of view of an ISP,
or a network operator, who owns the physical server infrastructure in the edge
network. So, given a certain network topology with servers located in every
node, we assume that arbitrary nodes of that topology have links to a third
party cloud. Then, in order to study the effect of migrations and replications
have in the network and on the operational costs, we need to define two
placement steps: one considering low traffic and one considering high traffic in
the network. In this way, with low traffic we use it as an initial placement of
VNFs where no replications are allowed, and of course no migrations have been
performed from any previous step, and with high traffic some of the VNFs need to
be migrated or replicated in order to provide service to the new traffic
demands. On the other hand, thanks to the flexibility that NFV provides, we
consider that the ISP is able to place VNFs in a third party cloud service
provider in case, for instance, the network or the servers usage needs to be
alleviated. In this case, this will incur additional costs for the ISP to pay
when deploying VMs or instantiating containers in the third-party cloud. From
the ISP point of view, neither the utilization of cloud servers nor the
utilization of links connecting to the cloud is relevant, but the costs of that
usage.

\newcommand{\nN}{n \in \mathbb{N}}
\newcommand{\mN}{m \in \mathbb{N}}
\newcommand{\nNp}{n \in \mathbb{N}_p}
\newcommand{\m}{m}
\newcommand{\mNp}{\m \in \mathbb{N}_p}
\newcommand{\y}{y}
\newcommand{\xX}{x \in \mathbb{X}}
\newcommand{\xXE}{x \in \mathbb{X}_E}
\newcommand{\xXC}{x \in \mathbb{X}_C}
\newcommand{\yX}{\y \in \mathbb{X}}
\newcommand{\xXn}{x \in \mathbb{X}_n}
\newcommand{\yXm}{\y \in \mathbb{X}_m}
\newcommand{\xXp}{x \in \mathbb{X}_p}
\newcommand{\lL}{\ell \in \mathbb{L}}
\newcommand{\pP}{p \in \mathbb{P}}
\newcommand{\pPs}{p \in \mathbb{P}_s}
\newcommand{\sS}{s \in \mathbb{S}}
\newcommand{\dD}{\lambda \in \Lambda}
\newcommand{\dDs}{\lambda \in \Lambda_s}
\newcommand{\ddDs}{\lambda' \in \Lambda_s}
\newcommand{\vVs}{v \in \mathbb{V}_s}
\newcommand{\tT}{t \in \mathbb{T}}
\newcommand{\yY}{y \in \mathbb{Y}}

\newcommand{\pl}{T_{p}^\ell}
\newcommand{\pnm}{T_{p}^{n, m}}
\newcommand{\loadratio}{\Gamma_{t(v)}^\mathrm{pro}}
\newcommand{\syncload}{\Gamma_{t(v)}^\mathrm{syn}}
\newcommand{\overhead}{\Theta_{t(v)}^s}
\newcommand{\overprovisioning}{\vartheta}
\newcommand{\replicable}{R_{t(v)}}
\newcommand{\Dl}{D_\ell}
\newcommand{\Dsmax}{D_s^{\mathrm{max}}}
\newcommand{\Dsmaxhat}{\hat{D}_{s}^\mathrm{{max}}}
\newcommand{\Dnet}{D^{\mathrm{net}}}
\newcommand{\processtrafficdelay}{D_{t(v)}^\mathrm{pro\_q}}
\newcommand{\processdelay}{D_{t(v)}^\mathrm{pro\_x}}
\newcommand{\maxdelay}{D_{t(v)}^\mathrm{pro,max}}
\newcommand{\minprocessdelay}{D_{t(v)}^\mathrm{pro\_x,min}}
\newcommand{\migrationdelay}{D^\mathrm{dwt}}
\newcommand{\Cx}{C_x^\mathrm{max}}
\newcommand{\Cl}{C_{\ell}^\mathrm{max}}
\newcommand{\maxcapserver}{C_{x, t(v)}^\mathrm{pro\_q,max}}
\newcommand{\energyidle}{E_{i}}
\newcommand{\energyutil}{ \alpha_{u}}
\newcommand{\serverothercosts}{K_x}
\newcommand{\functioncharges}{K_{t(v)}}
\newcommand{\penaltyparam}{\rho}

\newcommand{\xn}{x_n}
\newcommand{\rsp}{z_{p}^s}
\newcommand{\rspd}{z_{p}^{\lambda,s}}
\newcommand{\rspdd}{z_{p}^{\lambda',s}}
\newcommand{\fx}{f_x}
\newcommand{\fxsv}{f_x^{v,s}}
\newcommand{\fysv}{f_\y^{v,s}}
\newcommand{\Fxsv}{F_x^{v,s}}
\newcommand{\fxsvd}{f_{x,\lambda}^{v,s}}
\newcommand{\fxsvdd}{f_{x,\lambda'}^{v,s}}
\newcommand{\fysvd}{ f_{\y, \lambda}^{(v-1),s}}
\newcommand{\gxysv}{ g_{x, \y}^{v,s}}
\newcommand{\hsvp}{h_{p}^{v,s}}
\newcommand{\kl}{k_\ell}
\newcommand{\kx}{k_x}
\newcommand{\ksv}{k_v^s}
\newcommand{\penaltyvar}{q_p^{\lambda, s}}
\newcommand{\penaltyvarA}{\hat{q}_p^{\lambda, s}}
\newcommand{\penaltyvarB}{\hat{\hat{q}}_p^{\lambda, s}}
\newcommand{\penaltyvaraux}{y_p^{\lambda, s}}
\newcommand{\uell}{u_\ell}
\newcommand{\ux}{u_x}
\newcommand{\dsp}{d_p^{s}}
\newcommand{\dsdp}{d_p^{\lambda,s}}
\newcommand{\dsdph}{\hat{d}_p^{\lambda,s}}
\newcommand{\dpro}{d_{x, v, s}^{\mathrm{pro}}}
\newcommand{\dproq}{d_{x, v, s}^{\mathrm{pro\_q}}}
\newcommand{\dprox}{d_{x, v, s}^{\mathrm{pro\_x}}}
\newcommand{\dmgr}{d_{s}^{\mathrm{dwt}}}
\newcommand{\daux}{d_{x, \lambda}^{v,s}}
\newcommand{\trafficaux}{\Lambda_x^{v,s}}

\begin{table}
	\renewcommand{\arraystretch}{1.3}
	\caption{Notation}
	\label{notation}
	\centering
	\footnotesize
	\begin{tabular}{@{}c p{0.65\columnwidth}@{}}
		\toprule
		\textbf{Parameters}                   & \textbf{Meaning}                                                                               \\
		\midrule
		$\mathbb{N}$                          & set of nodes: $\mathbb{N} = \{1,...,N\}$, $\nN$.                                               \\
		$\mathbb{X}$                          & set of servers: $\mathbb{X} = \{1,...,X\}$, $\xX$.                                             \\
		$\mathbb{L}$                          & set of links: $\mathbb{L} = \{1,...,L\}$, $\lL$.                                               \\
		$\mathbb{P}$                          & set of all admissible paths: $\mathbb{P} = \{1,...,P\}$,
		$\pP$.                                                                                                                                 \\
		$\mathbb{S}$                          & set of SFCs: $\mathbb{S} = \{1,...,S\}$, $\sS$.                                                \\
		$\mathbb{T}$                          & set of VNF types: $\mathbb{T} = \{1,...,T\}$, $\tT$..                                          \\
		$\mathbb{V}_s$                        & ordered set of VNFs, where $\vVs$ is the
		$v$\textsuperscript{th} VNF in set $\mathbb{V}_s$.                                                                                     \\
		$\Lambda$                             & set of all traffic demands: $\Lambda = \{1,...,\Lambda\}$,
		$\dD$.                                                                                                                                 \\
		$\Lambda_s \subseteq \Lambda$         & subset of traffic demands $\dDs$ for SFC $\sS$.                                                \\
		$\mathbb{N}_p \subseteq \mathbb{N}$   & subset of ordered nodes
		traversed by path $\pP$.                                                                                                               \\
		$\mathbb{X}_n \subseteq \mathbb{X}$   & subset of servers attached
		to node $\nN$.                                                                                                                         \\
		$\mathbb{X}_p \subseteq \mathbb{X}$   & subset of ordered servers
		traversed by path $\pP$.                                                                                                               \\
		$\mathbb{X}_E \subseteq \mathbb{X}$   & subset of servers located at the edge.                                                         \\
		$\mathbb{X}_C \subseteq \mathbb{X}$   & subset of servers located at the cloud.                                                        \\
		$\mathbb{P}_s \subseteq \mathbb{P}$   & subset of admissible paths $\pPs$
		for SFC $\sS$.                                                                                                                         \\
		$\pl, \pnm$                           & binary, 1 if path $\pP$ traverses the link
		$\lL$ and 1 if path $\pP$ connects node $\nN$ and $\mN$ as source
		and destination path nodes, respectively.                                                                                              \\
		$\loadratio, \syncload$               & floats, load ratio of a VNF of
		type $t \in V_t$ and traffic ratio for synchronization traffic between
		two VNFs of type $t \in V_t$, respectively.                                                                                            \\
		$\overhead$                           & integer, VM overhead for VNF
		$\vVs$ of type $t \in V_t$.                                                                                                            \\
		$\replicable$                         & binary, 1 if VNF $\vVs$ of type $\tT$
		can be replicated.                                                                                                                     \\
		$\Cx, \Cl$                            & integers, maximum capacity of link $\lL$,
		maximum capacity of server $\xX$, respectively.                                                                                        \\
		$\maxcapserver$                       & integer, maximum processing capacity that can be assigned by a server $x$ to a VNF of type $t$ \\
		$\Dl, \Dnet$                          & floats, propagation delay of link $\lL$ and propagation delay that any demand can have in the network, respectively.                                                               \\
		$\Dsmax, \Dsmaxhat$                   & floats, maximum allowed delay for a SFC $\sS$ and maximum delay a SFC can have, respectively.  \\
		$\maxdelay$				              & floats, maximum allowed processing delay for a VNF of type $t$. 																					   \\
		$\processtrafficdelay, \processdelay$ & floats, delay of a VNF $v$ of type $t$ due to queues and processing, respectively.             \\
		$\migrationdelay$                     & float, service downtime duration caused by a migration.                                        \\
		\toprule
		\textbf{Variables}                    & \textbf{Meaning}                                                                               \\
		\midrule
		$\rsp$                                & binary, 1 if SFC $s$ is using path
		$\pPs$.                                                                                                                                \\
		$\rspd$                               & binary, 1 if traffic demand $\lambda$ from
		SFC $s$ is using path $\pPs$.                                                                                                          \\
		$\fx$                                 & binary, 1 if server $x$ is used to place at least one
		function, 0 otherwise.                                                                                                                 \\
		$\fxsv$                               & binary, 1 if VNF $\vVs$ from SFC $s$ is
		allocated at server $\xX$, 0 otherwise.                                                                                                \\
		$\fxsvd$                              & binary, 1 if VNF $\vVs$ from SFC $s$ is
		being used at server $\xX$ by traffic demand $\lambda$, 0 otherwise.                                                                   \\
		$\hsvp$                               & binary, 1 if VNF $\vVs$ from SFC $s$ is
		using path $\pP$ for state synchronization, 0 otherwise.                                                                               \\
		$\dsdp$                               & positive float, end-to-end delay perceived by a traffic
		demand $\lambda$ using path $p$.                                                                                                       \\
		$\penaltyvar$                         & positive float, penalty cost due to a traffic demand $\dDs$ from
		SFC $\sS$ using path $\pPs$.                                                                                                           \\
		$\uell$, $\ux$                        & positive floats, utilizations of a link $\lL$ and
		server $\xX$, respectively.                                                                                                            \\
		\bottomrule
	\end{tabular}
\end{table}

\section{Problem Formulation} \label{LP_models}

We model the network as $\mathbb{G}=(\mathbb{N} \cup \mathbb{X}, \mathbb{L})$
where $\mathbb{N} = \{1,...,N\}$ is a set of nodes, $\mathbb{X} = \{1,...,X\}$
is a set of servers and $\mathbb{L} = \{1,...,L\}$ is a set of directed links.
Specifically, $\mathbb{X}_n$ is a subset of servers $\xX$ attached to node
$\nN$. We denote the set of all SFCs as $\mathbb{S} = \{1,...,S\}$, where a
specific SFC $\sS$ is an ordered set of VNFs $\mathbb{V}_s = \{1,...,V_s\}$,
each VNF being of type $t$, $\tT$, $\mathbb{T} = \{1,...,T\}$, where $\vVs$ is
the $v$\textsuperscript{th} VNF in set $V_s$. Table \ref{notation} summarizes
the notations. It should be noted that the model is written such that it can be
efficiently used in optimization solvers. For instance, the big M method is
avoided when possible or its value is minimized in order to avoid numerical
issues with the solver.

\subsection{Objective Function}

We define the joint optimization problem as the minimization of the sum of three
different monetary costs: (i) costs of maintaining the servers, (ii) charges
applied when placing VNFs in third party clouds and (iii) penalty costs incurred
when violating SLA agreements due to the maximum service delay is exceed, i.e.,
\begin{equation} \label{total-costs}
	\text{\emph{minimize:}} \quad \sum_{\xXE} \kx^E + \sum_{\xXC} \kx^C + \sum_{\sS} \sum_{\dDs} \sum_{\pPs} \penaltyvar
\end{equation}
, where the $\kx^E$ is the operational cost of a server $x$ at the edge network,
$\kx^C$ are the charges of allocating VNFs into a server $x$ of a third party
cloud provider and $\penaltyvar$ is the penalty costs applied when the maximum
end-to-end delay of a SFC $\sS$ has been exceed for traffic demand $\dDs$
using path $\pPs$. We next specify each cost individually.

\subsubsection{Edge servers OPEX} To calculate the operational cost of each server
$\kx^E$ at the edge, we consider the following linear function:
\begin{equation}  \label{opex-servers}
	\forall \xXE: \kx^E = \fx \cdot \energyidle + \energyutil \cdot \ux + \serverothercosts
\end{equation}
, where the variable $\fx$ determines if a server $x$ is used or not,
$\energyidle$ is the energy consumption when the server is running idle,
$\energyutil$ determines the costs in relation to the server utilization $\ux$
and $\serverothercosts$ are the extra maintenance costs of a server
independently of the energy consumption which we remove from our analysis since
it has no effect on the optimization.

\subsubsection{Cost of Third-Party Clouds} To determine the charges $\kx^C$ for
all VNFs deployed in the cloud servers, we define:
\begin{equation} \label{functions-charges}
	\forall  \xXC: \kx^C = \sum_{\sS} \sum_{\vVs}  \functioncharges \cdot \fxsv
\end{equation}
, where the parameter $\functioncharges$ specifies the charges of a specific VNF
of type $t$, as long as the variable $\fxsv$ specifies that the VNF $v$ from
SFC $s$ has been allocated in a server $\xXC$ in the cloud. The parameter
$\functioncharges$ will vary depending on whether the VNF runs over a VM or as a
container.

\subsubsection{Penalty costs} If a certain traffic demand $\dDs$ from a SFC
$\sS$ using a path $\pPs$ exceeds the maximum service delay, then a penalty cost
defined by the positive float variable $\penaltyvar$ is applied, following:
\begin{subequations} \label{sla-penalty}
	\begin{equation}
		\forall \sS, \forall \pPs, \forall \dDs:
		\penaltyvar \geq \Bigg( \frac{\dsdp}{\Dsmax} - 1 \Bigg) \penaltyparam_s \cdot \rspd
	\end{equation}
	\begin{equation}
		\forall \sS: \Dsmax = \sum_{\vVs} \maxdelay + \Dnet
	\end{equation}
	\begin{equation}
		\forall \sS: \penaltyparam_s = \penaltyparam \sum_{\vVs}  \functioncharges
	\end{equation}
\end{subequations}
, where $\dsdp$ is the delay variable associated to a specific traffic demand
$\lambda$ using path $p$ (later explained in detail in Section
\ref{service_delay}), $\Dsmax$ is the maximum allowed end-to-end delay for a
specific SFC $s$, $\penaltyparam_s$ is the monetary penalty cost to be paid for
that service if $\dsdp$ is longer than $\Dsmax$ and $\rspd$ is the variable that
determines whether the traffic demand $\lambda$ from the SFC $s$ is using path
$p$ or not. $\Dsmax$ is calculated considering the maximum allowed processing
delay $\maxdelay$ for every VNF in the chain plus the maximum propagation delay
that a traffic demand can experience in the network $\Dnet$. The monetary
penalty costs $\penaltyparam_s$ is calculated as a percentage $\penaltyparam$ of
the selling price for that SFC $s$, which is determined by the sum of all
selling prices for every VNF in the chain. As selling prices for every VNF, we
use the same values $\functioncharges$ as the ones used for the cloud charges.
Due to the product of $\dsdp$ and $\rspd$, we use the following linearization
equations:
\begin{subequations} \label{sla-penalty-2}
	\begin{align}
		\forall \sS, \forall \pPs, \forall \dDs: \penaltyvaraux & \leq \dsdp                                \\
		\penaltyvaraux                                          & \leq \Dsmaxhat \cdot \rspd         \\
		\penaltyvaraux                                          & \geq \Dsmaxhat (\rspd - 1) + \dsdp
	\end{align}
\end{subequations}
, where $\penaltyvaraux$ has value $\dsdp$ when $\rspd$ is 1, 0 otherwise, and
$\Dsmaxhat$ is the worst possible case of service delay that a service can have
considering all service downtimes. So, $\Dsmaxhat$ is:
\begin{equation}
	\Dsmaxhat = \Dsmax + |\mathbb{V}_s| \cdot \migrationdelay
\end{equation}
, where $|\mathbb{V}_s|$ is the total number of VNFs in the SFC and
$\migrationdelay$ is the service downtime duration caused by a single migration.

\subsection{General Constraints}
The general constraints are related to the traffic routing, the VNF placement
and the mapping between VNFs and paths.

\subsubsection{Routing}
For a given network, the input set $\pPs$ is the set of all pre-calculated paths
for SFC $s$. The binary variable $\rspd=1$ indicates, that a traffic demand
$\dDs$ of the SFC $s$ is using path  $\pPs$. The first routing constraint
specifies that each traffic demand $\dDs$ from SFC $\sS$ has to use only one
path $\pPs$, i.e.:
\begin{equation} \label{onePathPerDemand}
	\forall \sS, \forall \dDs: \sum_{\pPs} \rspd = 1
\end{equation}
Then, the next constraint takes the activated paths from the variable $\rspd$
and activates the path for a certain SFC $s$:
\begin{equation} \label{activatePathForService}
	\forall \sS, \forall \pPs, \forall \dDs:  \rspd \leq \rsp \leq \sum_{\ddDs} \rspdd
\end{equation}
This forces  $\rsp$ to be 1 when at least one traffic demand is using path $p$,
whereas the right side forces to $\rsp$ to be 0 when no traffic demand $\lambda$
is using path $p$.

\subsubsection{VNF  placement}
VNF placement is modeled using the binary variable $\fxsvd$, which has only
value 1, if VNF $v$ from SFC $s$ is allocated at server $\xX$ and used
by traffic demand $ \dDs$. Similarly to (\ref{onePathPerDemand}), the next
constraint defines that each traffic demand $\dDs$ from SFC $\sS$ has to
traverse every VNF $\vVs$ in only one server $\xX$:
\begin{equation} \label{oneFunctionPerDemand}
	\forall \sS, \forall \vVs, \forall \dDs: \sum_{\xX} \fxsvd = 1
\end{equation}

Then, similarly to (\ref{activatePathForService}), the next constraint takes the
activated VNFs for each traffic demand from the variable $\fxsvd$ and activates
the VNF for a certain SFC $s$ as follows:
\begin{equation} \label{mappingFunctionsWithDemands}
	\forall \sS, \forall \vVs, \forall \xX, \forall \dDs: \fxsvd \leq \fxsv \leq  \!\!\!\! \sum_{\ddDs} \fxsvdd
\end{equation}
, where the left side forces to $\fxsv$ to be 1 when at least one traffic demand
$\dDs$ is using VNF $\vVs$ at server $\xX$ and the right side forces to $\fxsv$
to be 0 when no traffic demand is using that specific VNF $v$ on server $x$.
Likewise, we determine if a server is being used or not by constraining the
variable $\fx$ as:
\begin{equation}  \label{used-server}
	\forall \xX: \frac{1}{|\mathbb{S}||\mathbb{V}_s|} \sum_{\sS} \sum_{\vVs} \fxsv \leq \fx \leq \sum_{\sS} \sum_{\vVs} \fxsv     \text{  ,}
\end{equation}
where $\fx$ is 1 if at least one VNF from any SFC is allocated at server $\xX$,
0 otherwise.

\subsubsection{Mapping VNFs to paths}
The next equation maps the activated VNF to the activated paths defined in the
previous constraints. The first one defines how many times a VNF can be
replicated:
\begin{equation} \label{pathsConstrainedByFunctions}
	\forall \sS, \forall \vVs:  \sum_{\xX} \fxsv  \leq \replicable \sum_{\pPs} \rsp + 1 - \replicable
\end{equation}
, where $\replicable$ specifies if a certain VNF $v$ of type $t$ is replicable.
When $\replicable$ is 0, the total number of activated VNFs $\vVs$ from SFC
$\sS$ is $\sum_{\xX} \fxsv \leq 1$. In case the VNF is replicable, then the
maximum number of replicas is limited by the total number of activated paths
$\sum_{\pPs} \rsp$ for that specific SFC $s$. The next constraint activates
the VNFs on the activated paths:
\begin{equation}  \label{functionPlacement}
	\forall \sS, \forall \pPs, \forall \dDs , \forall \vVs: \rspd \leq \sum_{\xXp} \fxsvd
\end{equation}
If the variable $\rspd$ is activated, then every VNF $\vVs$ from SFC $\sS$
has to be activated in some server $\xXp$ from the path $\pP$ for a specific
traffic demand $\lambda$. When $\rspd$ is deactivated, then no VNFs can be
placed for that specific traffic demand. The last general constraint controls
that all VNFs $V_s$ from a specific SFC $s$ are traversed by every traffic
demand $\dDs$ in the given order, i.e.:
\begin{multline} \label{functionSequenceOrder}
	\forall \sS, \forall \dDs, \forall \pPs, \forall \vVs, \forall \nNp, \forall \mNp: \\
	\Bigg( \sum_{\m = 1}^{n} \sum_{\yXm}  \!\!  \fysvd \Bigg) \! - \! \!\!  \sum_{\xXn} \!\!  \fxsvd \geq \rspd \! - 1  \text{  for  }   1< v \leq |\mathbb{V}_s |
\end{multline}
, where the variable $\rspd$ activates the ordering constraint side (left side)
when is 1 and deactivates it, otherwise. Then, if path $\pP$ is activated, the
ordering is checked for every traffic demand $\dDs$ individually by using the
variable $\fxsvd$. Hence, for every traffic demand $\lambda$ of SFC $s$, the
$v$\textsuperscript{th} VNF is allocated at server $\xXn$ only if the previous
$(v - 1)$\textsuperscript{th} VNF is allocated at any server $\yXm$, where $m$
is the i\textsuperscript{th} node from 1 until $n$ traversed by path $p$. It
should be noted, that the correct sequence of VNFs relies on the correct
sequence of subset of servers, i.e.  $\xXn$. This assumes that the  correct
sequence of VNFs inside these subsets  is organized by the local routing, which
may be located at the node $n$ or at a local switch not modeled in detail.

\subsection{Traffic and Performance Constraints}

\subsubsection{Initial placement parameter}

Since the optimization process follows two different phases, after the initial
placement we take the value of variables $\fxsv$ and convert them into the input
parameters $\Fxsv$ for the next next placement step, i.e.
\begin{equation} \label{initialsolutionmapping}
	\forall \sS, \forall \vVs, \forall \xX:      \fxsv   \Rightarrow  \Fxsv
\end{equation}
The parameter $\Fxsv $ determines if a VNF $v$ of a service chain $s$ was placed
on server $x$ during the initial placement.

\subsubsection{Migration and replications}

To identify a migration, the initial solution has to be mapped using
\eqref{initialsolutionmapping}. For evaluation and comparison reasons, the
number of migrations $n_{mgr}$ can be calculated as follows:
\begin{equation} \label{num_mgr}
	n_{mgr} = \sum_{\xX} \sum_{\sS} \sum_{\vVs} \Fxsv (1 - \fxsv)
\end{equation}
, where the variable $\Fxsv$ specifies the initial placement, and $\fxsv$
indicates if the same VNF has been removed from the same server $x$. On the
other hand, to count the number of replications for every VNF, we use:
\begin{equation} \label{num_rep}
	n_{rep} = \sum_{\sS} \sum_{\vVs} \big[ (\sum_{\xX} \fxsv) - 1 \big]   \text{ ,}
\end{equation}
where we do account for the original VNF.

\subsubsection{Synchronization traffic}

When performing replications of a specific VNF, the statefulness between the
original and replicas has to be maintained in order to be reliable against VNF
failures and avoiding the lost of information. For this reason, we consider that
when a VNF is replicated, the generated synchronization traffic between replicas
and the original has to be also considered. The amount of the state
synchronization traffic depends on the state space and its time dynamic, where
it is assumed, that each VNF has full knowledge on the state of all its
instances used to implement the VNF $\vVs$. Let us assume, that this amount is
proportional to the total traffic offered to the SFC weighted by an
synchronization ratio $\syncload$, which depends on the type of VNF $t$. In
summary, the directional traffic from a VNF to its replica is given by
$\syncload  | \Lambda_s | $, and its routing should be optimized within the
network.

In order to know if the same VNF $\vVs$ from SFC $s$ is placed in two
different servers $\xX$ and $\yX$, we define:
\begin{equation} \label{sync-traffic}
	\forall \sS, \forall \vVs, \forall \xX, \forall \yX, y \! \neq \! x:  \gxysv = \fxsv \fysv
\end{equation}
, where the variable $\gxysv$ is 1 only when both variables $\fxsv$ and $\fysv$
are also 1, and 0 otherwise. In this way, this variable is used to know if two
different servers have the same VNF placed, which means that model is allocating
one replica. We use the well-known linearization method when multiplying two
binary variables. In case $\gxysv = 1$, we need to carry the synchronization
traffic from server $x$ to $y$, by selecting only one predefined path  between
them, i.e.:
\begin{multline} \label{hpvs}
	\forall \sS, \forall \vVs,  \forall n, m \in \mathbb{N},  n \neq m, \forall \xXn, \forall \yXm:  \\
	\gxysv  \leq  \sum_{\pP} \hsvp \cdot \pnm \leq 1     \quad    \text { ,}
\end{multline}
\begin{multline} \label{hpvs_2}
	\forall \sS, \forall \vVs,  \forall n, m \in \mathbb{N},  n \neq m:  \\
	\sum_{\pP} \hsvp \cdot \pnm  \leq  \sum_{\xXn} \sum_{\yXm}  \gxysv  \quad    \text { ,}
\end{multline}
, where the constant $ \pnm = 1$ indicates, that the path $ \pP$ exists which
connects servers  $ \xXn$ and $ y \in X_m$ using the shortest path between nodes
$n$ and $m$. The right term of \eqref{hpvs} guarantees that only one path $\pP$
is selected by variable $\hsvp$. Moreover, \eqref{hpvs_2} guarantees that this
path is only used if at least one $\gxysv$ is 1. Note that $\hsvp$ is a binary
variable used for every VNF $v$ of SFC $s$.

\subsubsection{Link and server utilization}

The utilization of a link is calculated as follows:
\begin{multline} \label{linkutil}
	\forall \lL: \uell =   \frac{1}{\Cl}  \sum_{\sS} \sum_{\pPs} \sum_{\dDs} \lambda \cdot \pl \cdot \rspd +  \\
	\frac{1}{\Cl}  \sum_{\pP}  \pl \sum_{\sS} \sum_{\vVs}  \syncload \cdot  |\Lambda_s| \cdot \hsvp  \leq 1  \text{ ,}
\end{multline}
where $\lambda \cdot \pl$ adds the traffic demands from SFC $\sS$ when a path
$\pPs$ traverses the link $\lL$. Then, the variable $\rspd$ specifies if the
traffic demand $\lambda$ from SFC $s$ is using path $p$. The second term is the
sum of the extra traffic generated due to the state synchronization between VNFs
$\vVs$ from SFC $s$, which is proportional to its total traffic $|\Lambda_s|$
multiplied by the synchronization traffic ratio $\syncload$ of the VNF of type
$t$. This traffic is only added, if the variable $\hsvp$ is 1, which indicates
that path $\pP$ is used for synchronization by a VNF $v$ from SFC $s$, and the
link  $ \lL$ belongs to this path. Both summation terms are divided by the
maximum link capacity $\Cl$ to restrict the utilization.

The processing load of a server is derived as
\begin{equation} \label{server_load}
	\gamma_x = \sum_{\sS} \sum_{\vVs} \Big(  \loadratio  \sum_{\dDs} \lambda \cdot \fxsvd + \overhead \cdot \fxsv \Big)
\end{equation}
, where the first term sums the traffic $\dDs$ that is using the VNF $\vVs$ from
SFC $\sS$ at server $\xX$, which is determined by the variable $\fxsvd$, and
multiplied by the processing load ratio $\loadratio$ of the VNF of type $t$. The
second term adds the overhead generated by the VM where the VNF is running and
is only added, when the variable $\fxsv$ determines that this VNF is placed in
server $x$. That term will be omitted when the VNF is deployed as \ctvnf,
instead. Then, the utilization follows to be given by
\begin{equation} \label{serverutil}
	\forall \xX:  \ux  =    \frac{\gamma_x}{\Cx}      \leq 1    \text{ ,}
\end{equation}
where  $\Cx$  is  the maximum processing capacity.

\subsubsection{Service delay} \label{service_delay} Since every service has a
maximum allowed delay $\Dsmax$ specified in the SLA agreement, in case of
exceeding it, some penalty costs are applied. In our model, and for simplicity,
we take into account the propagation delay due to the traversed links, the
processing delay that every VNF requires in the servers and, where applicable,
the downtime delays caused by the interruption of the service during the
migrations of VNFs.

\emph{Processing delay}: The processing delay $\dpro$ of a VNF $v$ in a server
$x$ depends, on the one side, on the amount of traffic being processed by a
specific VNF, described by $\dproq$, and on  $\dprox$, which is related to the
VNF type and  the total  server  load $\ux$, given as
\begin{subequations} \label{processing_delay_equations}
	\begin{equation}
		\forall \sS, \forall \vVs, \forall \xXp:  \dpro = \dproq + \dprox
	\end{equation}
	\begin{equation}
		\dproq = \processtrafficdelay \frac{ \loadratio \cdot \sum_{\dDs} \fxsvd \cdot \lambda}{\maxcapserver} \label{processing_delay_equations_B}
	\end{equation}
	\begin{equation}
		\dprox =  \minprocessdelay \cdot \fxsv + \processdelay \cdot \ux  \label{processing_delay_equations_C}
	\end{equation}
\end{subequations}

In \eqref{processing_delay_equations_B}, the numerator of $\dproq$  determines
the total processing load assigned to the VNF of type $t$, which is controlled
by the variables $\fxsvd$. Thus, if the assigned processing load is equal to
$\maxcapserver$, the VNF adds the processing delay $\processtrafficdelay$. The
second delay term, given in \eqref{processing_delay_equations_C}, adds the load
independent minimum delay associated to the usage of a type of this VNF, and a
delay part which increases with the server utilization. As a consequence the
processing delay $\dpro(\vec{ \lambda})$ depends on the server $x$, the used VNF
type and linearly increases with increasing traffic. Furthermore, the dependency
on all traffic demands is denoted by the vector $\vec{ \lambda}$, which is
omitted for simplicity in \eqref{processing_delay_equations}.

\emph{Downtime duration}: If a VNF $v$ of SFC $s$ has to be migrated, we assume
an interruption of the service with duration $\migrationdelay$. Thus, the total
service downtime will consider the migration of all VNFs in that SFC which
yields a constraint as follows:
\begin{equation}  \label{migration_delay_equations}
	\forall \sS: \dmgr = \migrationdelay \sum_{\xX} \sum_{\vVs} \Fxsv (1 - \fxsv)
\end{equation}
, where the parameter $\Fxsv$ determines if a VNF $v$ was placed on server $x$
during the initial placement. Thus, if a VNF migrates to another server $y \neq
	x$, the variable $\fxsv$ is equal to zero and the service downtime
$\migrationdelay$ has to be taken into account.

\emph{Total delay}: Because the model allows that different traffic demands per
service can be assigned to different paths, we define individual end-to-end
delay $\dsdph$ for every traffic demand, as follows:
\begin{multline} \label{exact_service_delay}
	\forall \sS, \forall \dDs, \forall \pPs: \\
	\dsdph =  \sum_{\lL} \Dl \cdot \pl  + \sum_{\xXp} \sum_{\vVs}    \dpro(\vec{\lambda}) \cdot  \fxsvd   + \dmgr
\end{multline}
The first term is the propagation delay, where $\Dl$ is the delay of the link
$\ell$, and $\pl$ specifies if the link $\ell$ is traverses by path $\pPs$. The
second term adds the processing delays caused by all VNFs from the SFC placed on
the servers $\xXp$, in which the variable $\fxsvd$ has to ensure that the demand
$\lambda$ is processed at an specific server $x$. Finally, the third term is the
total downtime duration due to the migrations of that service chain. It should
be noted that the second term of \eqref{exact_service_delay} includes a
nonlinear relation between the binary variable $f^{v,s}_{x, \lambda}$ and the
delay variable $\dpro$, which also depends on all decision variables
$f^{v{'},s{'}}_{x, \lambda{'}}$. To solve that, we introduce a new delay
variable $\daux$, which  is  bounded as follows:
\begin{equation} \label{new-variable}
	\dpro - \maxdelay(1 - \fxsvd)  \leq  \daux   \leq \maxdelay \cdot \fxsvd
\end{equation}
If the VNF is selected at server $x$ by $\fxsvd=1$, the variable is lower
bounded by the exact delay $\dpro$ and upper bounded by the maximum VNF delay
$\maxdelay$. Since $\dpro \leq \daux \leq \maxdelay$, the specific delay of a
VNF can be restricted. If the VNF is not selected, i.e., $\fxsvd=0$, the
variable has value $\daux=0$, since the constant $\maxdelay$ makes the left size
of \eqref{new-variable} to be negative. Hence, the end-to-end delay is mapped to
an upper and lower bounded variable $\dsdp$ given as
\begin{multline} \label{total_service_delay}
	\forall \sS, \forall \dDs, \forall \pPs: \\
	\dsdp =  \sum_{\lL} \Dl \cdot \pl  + \sum_{\xXp} \sum_{\vVs} \daux + \dmgr \quad \text{,}
\end{multline}
in which the bounding feature is used in the optimization scenarios described
next.

\subsection{Optimization Scenarios}
According to Section \ref{system_model}, we divide the placement into two
different phases, one where the VNFs are placed in the network only considering
low traffic and another one when some of these VNFs are migrated and/or
replicated in order to serve the new traffic demands. For the initial placement,
from the set of traffic demands $\Lambda_s$ of SFC $\sS$, only a subset of
demands $\dDs$ is selected, where each demand is randomly selected with
probability $R$ to be an element of the reduced set $\Lambda_s'$. If no demand
is selected, the reduced set contains at least one demand randomly selected out
of $\Lambda_s$. For both cases, initial placement and second placement, the
constraints \eqref{onePathPerDemand} - \eqref{functionSequenceOrder} and
\eqref{linkutil} - \eqref{total_service_delay} apply. Only for the second
placement, the constraints \eqref{initialsolutionmapping}, \eqref{sync-traffic}
- \eqref{hpvs_2} also apply.

\section{Online Heuristic Approaches}

Since the model presented is a MILP optimization problem and these models are
known to be NP-hard \cite{Bulut2015}, in this section we propose a greedy
algorithm to work as an online solution and, First-Fit and Random-Fit algorithms
for comparison purposes.

\subsection{First-Fit and Random-Fit algorithms}

Both \emph{First-Fit} (FF) and \emph{Random-Fit} (RF) algorithms pseudo code are
described in Algorithm \ref{algorithm:ff_rf}. While both approaches share most
of the code, the variable \emph{alg} (line \ref{simple_placement_procedure})
specifies whether the code has to run FF or RF. The process starts with a loop
where every demand from every SFC is going to be considered (line
\ref{loop_simple}). The first step is to then retrieve all the paths with enough
link resources to assign traffic demand $\lambda$ and that also connect both
source and destination nodes (line \ref{get_admissible_paths_simple}). These
paths are saved into $\mathbb{P}_s'$, from where one admissible path $p$, first
one for FF or a random one for RF, is selected (line \ref{choose_path_simple}).
In this point, we make sure here that in this path, there are enough server
resources to allocate all the VNFs for SFC $s$ (line
\ref{for_functions_simple}). From that path, we start selecting a server for
every VNF $v$ from SFC $s$. First, we retrieve all servers with enough free
capacity to allocate the VNF $v$ and to provide service to demand $\lambda$
(line \ref{get_available_servers_simple}), and then we choose the first
available server in FF or a random one in RF (line \ref{choose_server_simple}).
It is to be noted here, that to satisfy VNF ordering (see equation
\ref{functionSequenceOrder}), the procedure \emph{chooseServer} will return a
valid server from before/after the previous/next VNF allocated. While for the FF
case, we assure in line \ref{choose_path_simple} that there will always be a
server where to allocate the next VNF in the chain, in RF case we make sure here
(line \ref{choose_server_simple}) that after the random server selected there is
still place to allocate all the rest of the VNFs from the chain in next servers
in the path, or we select another server instead. In line
\ref{add_function_to_server_simple}, we assign the demand and the VNF to the
server (i.e. equations (\ref{oneFunctionPerDemand}) and
(\ref{mappingFunctionsWithDemands})). After all the VNFs have been placed, the
next step is to route traffic demand $\lambda$ to path $p$ (line
\ref{route_demand_to_path_simple}), to finally add the synchronization traffic
for the service chain (line \ref{add_synch_traffic_simple}).

\begin{algorithm}[!t]
	\caption{First-Fit and Random-Fit algorithms}
	\begin{algorithmic}[1]
		\Procedure{simplePlacement}{$alg$} \label{simple_placement_procedure}
		\For{$\sS$, $\dDs$} \label{loop_simple}
		\State $\mathbb{P}_s' \gets$ getAdmissiblePaths($s$, $\lambda$) \label{get_admissible_paths_simple}
		\State $p \gets$ choosePath($alg$, $\mathbb{P}_s'$) \label{choose_path_simple}
		\For{$\vVs$} \label{for_functions_simple}
		\State $\mathbb{X}_p' \gets$ getAvailableServers($s$, $\lambda$, $v$, $p$) \label{get_available_servers_simple}
		\State $x \gets$ chooseServer($alg$, $\mathbb{X}_p'$) \label{choose_server_simple}
		\State addVNFToServer($s$, $v$, $\lambda$, $x$) \label{add_function_to_server_simple}
		\EndFor
		\State routeDemandToPath($s$, $p$, $\lambda$) \label{route_demand_to_path_simple}
		\State addSynchronizationTraffic($s$)  \label{add_synch_traffic_simple}
		\EndFor
		\EndProcedure
	\end{algorithmic}
	\label{algorithm:ff_rf}
\end{algorithm}

\begin{algorithm}[!t]
	\caption{Greedy algorithm}
	\begin{algorithmic}[1]
		\Procedure{main}{}
		\For{$\sS$, $\lambda \in \Lambda'_s$}  \label{for_init_demands}
		\State allocateDemand($s$, $\lambda$) \Comment{go to \ref{allocate_demand}}
		\EndFor
		\For{$\sS$}
		\For{$\lambda \in \Lambda''_s$} \label{for_rest_demands}
		\State allocateDemand($s$, $\lambda$) \Comment{go to \ref{allocate_demand}}
		\EndFor
		\State addSynchronizationTraffic($s$)  \label{add_synch_traffic}
		\EndFor
		\State $b \gets$ computeObjVal() \label{compute_obj_val}
		\State findNewIncumbent($b$) \label{find_new_incumbent}  \Comment{go to \ref{find_new_incumbent_procedure}}
		\EndProcedure
		\Procedure{findNewIncumbent}{$b$} \label{find_new_incumbent_procedure}
		\For{$s \in S'$}  \label{reallocation_start}
		\For{$s \in S'$}
		\For{$\dDs$}
		\State removeDemandFromVNFs($s$, $\lambda$) \label{remove_demand_functions}
		\State removeDemandFromPath($s$, $\lambda$) \label{remove_demand_path}
		\State $b' \gets$ simplePlacement($RF$) \label{simple_placement}
		\If{$b' < b$} \label{if_better_value}
		\State $b \gets b'$
		\State setNewIncumbent()
		\Else \State undoPlacement() \label{end_better_value}
		\EndIf
		\EndFor
		\State reAssignSyncTraffic($s$) \label{reassign_synch_traffic}
		\EndFor
		\EndFor
		\EndProcedure
		\Procedure{allocateDemand}{$s$, $\lambda$} \label{allocate_demand}
		\State $\mathbb{P}_s' \gets$ getAdmissiblePaths($s$, $\lambda$) \label{get_admissible_paths}
		\For{$p \in \mathbb{P}_s'$}
		\State $p \gets$ choosePath($s$, $\lambda$, $\mathbb{P}_s'$) \label{choose_path_p} \Comment{go to \ref{choose_path}}
		\For{$\vVs$}
		\State $\mathbb{X}_p' \gets$ getAvailableServers($s$, $\lambda$, $p$, $v$) \label{get_available_servers}
		\State $x \gets$ chooseServer($s$, $\lambda$, $p$, $v$, $\mathbb{X}_p'$) \label{choose_server_x} \Comment{go to \ref{choose_server}}
		\State addVNFToServer($v$ , $x$) \label{add_function_to_server}
		\EndFor
		\State routeDemandToPath($s$, $p$, $\lambda$) \label{route_demand_to_path}
		\EndFor
		\EndProcedure
		\Procedure{choosePath}{$s$, $\lambda$, $\mathbb{P}_s'$} \label{choose_path}
		\State $p \gets$ getUsedPathForDemandInitPlacement($s$, $\lambda$, $\mathbb{P}_s'$) \label{get_used_path_for_demand_init}
		\If{$p$}
		return $p$
		\EndIf
		\State $p \gets$ getUsedPathInitialPlacement($s$, $\mathbb{P}_s'$) \label{get_used_path_init}
		\If{$p$}
		return $p$
		\EndIf
		\State $p \gets$ getUsedPathForSFC($s$, $\mathbb{P}_s'$) \label{get_used_path_for_service}
		\If{$p$}
		return $p$
		\EndIf
		\State return getPathWithShortestDelay($s$, $\lambda$, $\mathbb{P}_s'$) \label{get_path_shortest_delay}
		\EndProcedure
		\Procedure{chooseServer}{$s$, $\lambda$, $p$, $v$, $\mathbb{X}_p'$} \label{choose_server}
		\State $\mathbb{X}_p' \gets$ removeServersPreviousVNFs($\mathbb{X}_p'$) \label{remove_servers_previous_vnfs}
		\State $\mathbb{X}_p' \gets$ removeServersNextVNFs($\mathbb{X}_p'$) \label{remove_servers_next_vnfs}
		\State $x \gets$ getUsedServerDemandInitialPlace($s$, $v$, $\lambda$, $\mathbb{X}_p'$) \label{get_used_server_demand_init}
		\If{x}
		return x
		\EndIf
		\State $c \gets$ returnCloudServer($\mathbb{X}_p'$) \label{get_cloud_server}
		\State $x \gets$ getUsedServerInitialPlacement($s$, $v$, $\mathbb{X}_p'$) \label{get_used_server_init}
		\If{indexOf($x$) < indexOf($c$)}  \label{use_server_init}
		return x
		\EndIf
		\State $x \gets$ getUsedServerForSFC($s$, $v$, $\mathbb{X}_p'$) \label{get_used_server}
		\If{indexOf($x$) < indexOf($c$)} \label{use_server}
		return x
		\EndIf
		\State return $\mathbb{X}_p'$[0] \label{get_first_server}
		\EndProcedure
	\end{algorithmic}
	\label{algorithm:greedy}
\end{algorithm}

\subsection{Greedy algorithm}

The greedy algorithm pseudo code is described in Algorithm
\ref{algorithm:greedy}. The main procedure starts first allocating all demands
$\Lambda'_s$ from all SFCs used during the initial placement (line
\ref{for_init_demands}) and then, continues with the rest of traffic demands
$\Lambda''_s$ (line \ref{for_rest_demands}), where $\Lambda''_s = \Lambda_s
\setminus \Lambda'_s$. In this way, we assure there are no migrations of VNFs
during the initial placement. In both cases, we are pointing out to the function
defined in line \ref{allocate_demand} which allocate traffic demand in the
network while traversing the required VNFs (explained later in detail). After
the allocation of all demands is done, we add the synchronization traffic
between the replicas (line \ref{add_synch_traffic}). Once all the resources are
allocated in the network, we go back to line \ref{compute_obj_val} and save the
current objective value. From this point, starting from line
\ref{reallocation_start}, we go over all SFCs and try to improve the current
solution by randomly switching demands. To do that, for every traffic demand, we
first remove it from the current VNFs assigned and from the network (line
\ref{remove_demand_functions} and \ref{remove_demand_path}) and then perform a
random fit placement (line \ref{simple_placement}) following Algorithm
\ref{algorithm:ff_rf} using a random placement. If the new objective value $b'$
is lower, the placement is set as incumbent, otherwise is undone (lines
\ref{if_better_value}-\ref{end_better_value}). After the reallocation of VNFs,
the synchronization traffic is reassigned (line \ref{reassign_synch_traffic}).

Going into detail on the allocation of demands VNFs (line
\ref{allocate_demand}), the procedure starts by retrieving all paths with enough
free link resources in $\mathbb{P}_s'$ (line \ref{get_admissible_paths}). Then,
we choose a path $p$ inside of a loop from all retrieved paths (line
\ref{choose_path_p}, details explained later). This is done to make sure in case
a path cannot be used for allocating all VNFs, the algorithm tries with the next
one. Once the path is selected, we start with the placement of all VNFs on the
selected path. First, all the available servers for an specific VNF $v$ on path
$p$ are retrieved in variable $\mathbb{X}_p'$ (line
\ref{get_available_servers}), then we choose one server $x$ for that specific
VNF in line \ref{choose_server_x} (this procedure explained later) and place the
VNF (line \ref{add_function_to_server}). In case the VNF has been already placed
by another demand of the same service, the demand is associated to that VNF,
instead. Finally after all VNFs are placed, we route the demand over path (line
\ref{route_demand_to_path}).

When selecting a path for a specific traffic demand (line \ref{choose_path}), we
execute the following methods in this specific order: return the already used
path for the same demand $\lambda$ during the initial placement (line
\ref{get_used_path_for_demand_init}), return any used path for SFC $s$ during
the initial placement (line \ref{get_used_path_init}), return any used path for
SFC $s$ (line \ref{get_used_path_for_service}) or return the path with shortest
path delay (line \ref{get_path_shortest_delay}). If one method does not return a
path, then the next one is executed. On the other hand, when choosing a server
for a specific VNF (line \ref{choose_server}), we first remove servers in the
path that have already allocated VNFs before/after the current VNF in order to
satisfy with chain order equation (\ref{functionSequenceOrder}) in lines
\ref{remove_servers_previous_vnfs} and \ref{remove_servers_next_vnfs}. And then,
we are first try to use a server already used for VNF $v$ and demand $\lambda$
during the initial placement (line \ref{get_used_server_demand_init}). If none,
then we first return the cloud server, if any in path $p$, into variable $c$
(line \ref{get_cloud_server}) and then we try to get any server already used
during the initial placement for VNF $v$ into variable $x$ (line
\ref{get_used_server_init}). If that server is located in a node of the path
that comes before in the sequence order than the cloud node, then we use it
(line \ref{use_server_init}), otherwise, we try to use any other already used
server for that VNF also if it is before in the path than the cloud (lines
\ref{get_used_server} and \ref{use_server}). In case there is no cloud in the
path, both previous conditions are always true. If none of the previous methods
worked, then we just return the first available server (line
\ref{get_first_server}). This procedure is done in order to minimize the number
of migrations and replications that cause an increment of the monetary costs.

\begin{table*}[!t]
	\caption{Parameters for VM and CT based VNFs}
	\label{tab:vnf-parameters}
	\centering
	\renewcommand{\arraystretch}{1.2}
	\begin{tabular}{c|ccccccccc}
		\toprule
		   & $\overhead$ & $\loadratio$         & $\syncload$          & $\maxcapserver$     & $\processtrafficdelay$ & $\processdelay$       & $\minprocessdelay$    & $\maxdelay$            & $\functioncharges$ (cloud) \\[1.5ex]  \hline
		VM & 7           & \multirow{2}{*}{1.2} & \multirow{2}{*}{0.1} & \multirow{2}{*}{72} & \multirow{2}{*}{3 ms}  & \multirow{2}{*}{5 ms} & \multirow{2}{*}{2 ms} & \multirow{2}{*}{10 ms} & 0.0069 \$/h                \\\cline{1-2} \cline{10-10}
		CT & 0           &                      &                      &                     &                        &                       &                       &                        & 0.1199988 \$/h             \\
		\bottomrule
	\end{tabular}
\end{table*}

\subsubsection{Computational complexity}
In terms of complexity from bottom to top, for the procedure \emph{chooseServer}
in line \ref{choose_server} considering $V_{L}$ as the length of the longest
SFC, it is in the order of $O(cs) = O(V_{L} \cdot |\mathbb{X}|)$. The procedure
\emph{choosePath} in line \ref{choose_path} is in the order of $O(cp) =
	O(P_{S})$ where $P_{S}$ is the number of paths per SFC. The procedure
\emph{allocateDemand} in line \ref{allocate_demand} is calculated based on
$O(cs)$ and $O(cp)$, and is in the order of $O(ad) = O(P_{S} \cdot L_{P} [O(cp)
			+ V_{L} \cdot O(cs)])$. The \emph{addSynchronizationTraffic} procedure specified
in line \ref{add_synch_traffic} is in the order of $O(st) = O(V_{L} \cdot
	|\mathbb{X}^2| \cdot |\mathbb{P}|)$. The \emph{findNewIncumbent} procedure in
line \ref{find_new_incumbent} is in the order of $O(fi) = O(|\Lambda|^2)$ and
the \emph{reAssignSynTraffic} in line \ref{reassign_synch_traffic} is $O(rs) =
	V_{L} \cdot |\mathbb{X}^2| \cdot |\mathbb{P}|$. So, finally, the complexity of
the entire algorithm is in the order of $O(|\Lambda| \cdot O(ad) \cdot O(st) +
	O(fi) \cdot O(rs))$.


\section{Performance evaluation}

\begin{figure}[!t]
	\centering
	\subfloat[7 nodes network]{\includegraphics[width=0.35\columnwidth]{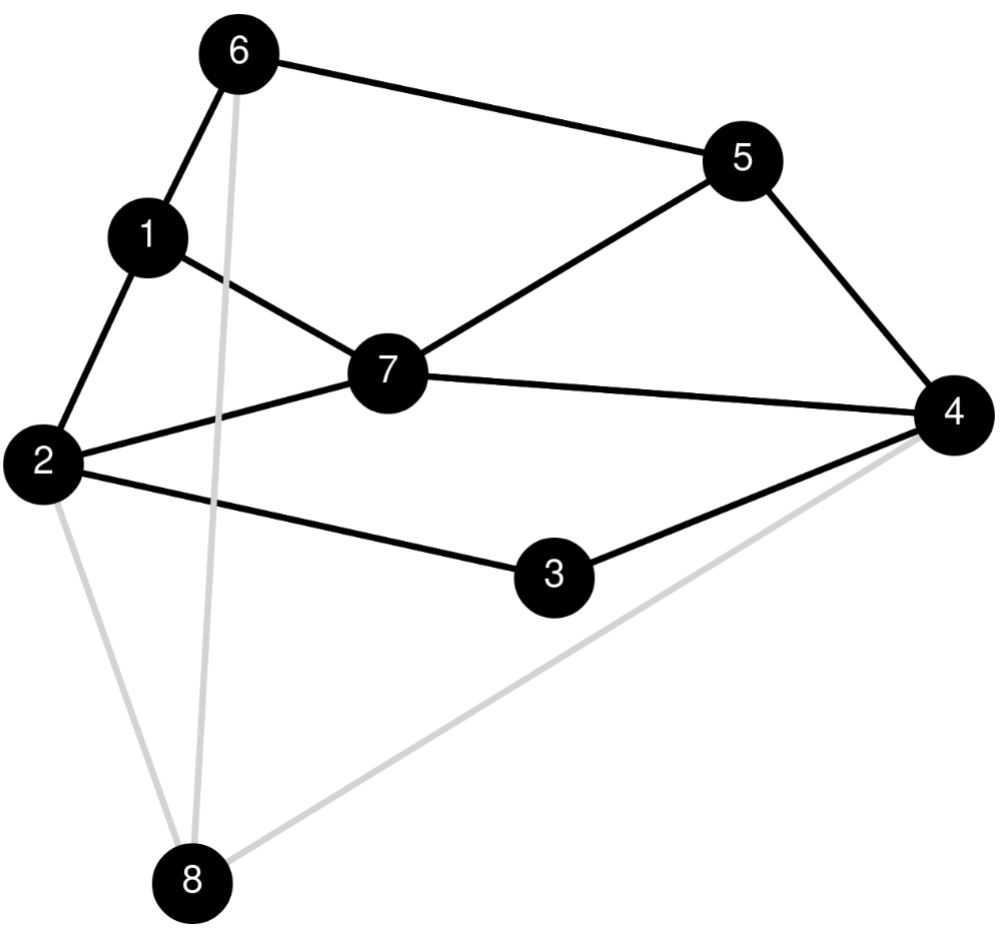}%
		\label{fig:small-network}}
	\hfil
	\subfloat[44 nodes network]{\includegraphics[width=0.55\columnwidth]{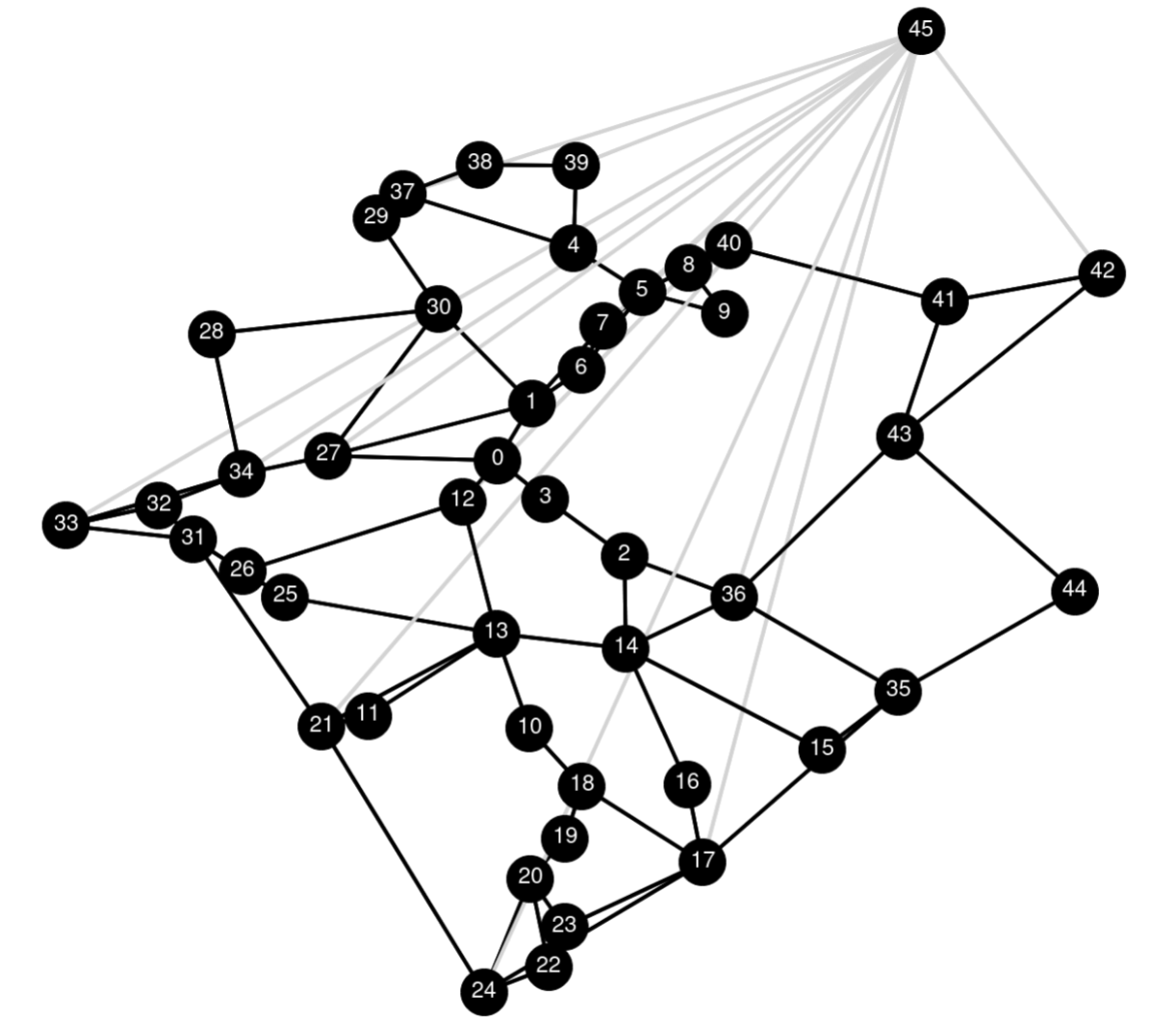}%
		\label{fig:large-network}}
	\caption{Network topologies used for the performance evaluation.}
\end{figure}

We first evaluate a smaller size network using the MILP model (implemented using
Gurobi Optimizer) and heuristics, and, then, we evaluate a larger network using
heuristics only. The small size network analyzed has 7 nodes and 20 directed
links with 500 units of capacity each (see Fig. \ref{fig:small-network}),
located within a small city area, with 3 nodes nodes are connected to a distant
cloud node (node \emph{c}). The second, large network has 44 nodes and 140
directed links with 5000 units of capacity (see Fig. \ref{fig:large-network})
with 13 nodes connected to a cloud node. To calculate the propagation delays in
real geographic locations, we used examples of Braunschweig (edge) and Frankfurt
(cloud) in Germany for the small network. For the larger network, which is based
of Palmetto network (South Carolina), we used the real locations whereas the
cloud is located in North Virginia in the USA. In both cases, the location of the
cloud is chosen based on the closest common locations that cloud providers
offer. We now calculate the propagation delay by calculating the distance
between nodes from their latitude and longitude using the Haversine method
divided by 2/3 the speed of light. We thereby assume the links used to connect
to the cloud are out of the ISP premises, have sufficient capacity for any
demand, and therefore do not impact the analysis. In all cases, the propagation
delay due to the distance to the cloud it will be taken into account. Every node
in the small network has one server, while in the large network every node has 8
servers, in both cases with all servers with 1000 units of capacity. On the
other hand, the cloud node has in both cases one server with a large capacity
also to not interfere on the results. In the edge network, the costs due to the
power consumption for every server specified in equation \eqref{opex-servers}
have been calculated considering the power consumption of a Dell PowerEdge R410
Rack in Watts server specified in the spec sheet \cite{Psu2011} multiplied by
the average monetary costs of the electricity in USA is 0.139 \$/kWh, so
$\energyidle$ = 0.0184453 and $\energyutil$ = 0.0095632. All other parameters
are the same for both networks.

\begin{figure*}[!t]
	\centering
	\subfloat[Total costs]{\includegraphics[width=0.50\columnwidth]{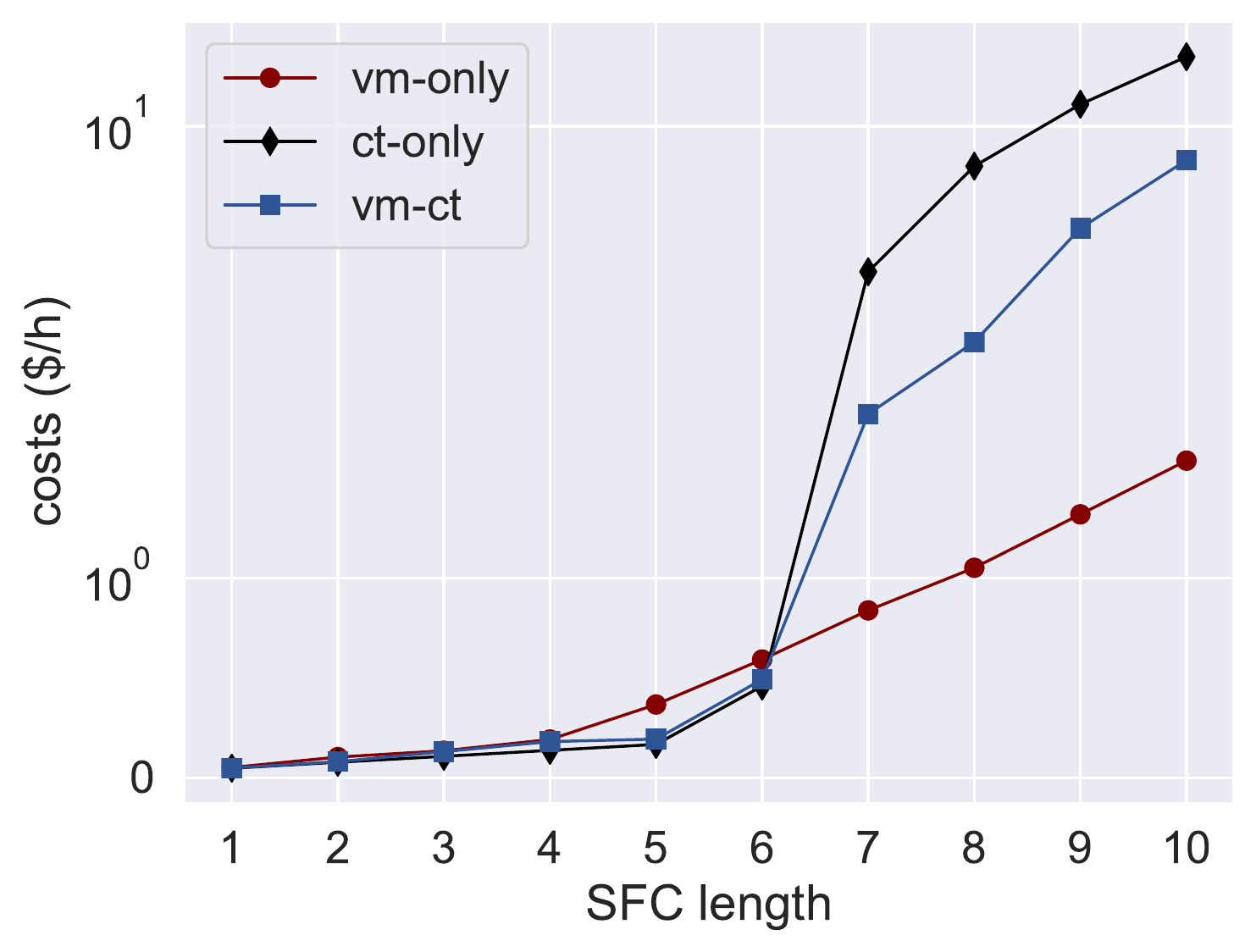}%
		\label{fig:7nodes_costs-sfc}}
	\hfil
	\subfloat[OPEX costs]{\includegraphics[width=0.50\columnwidth]{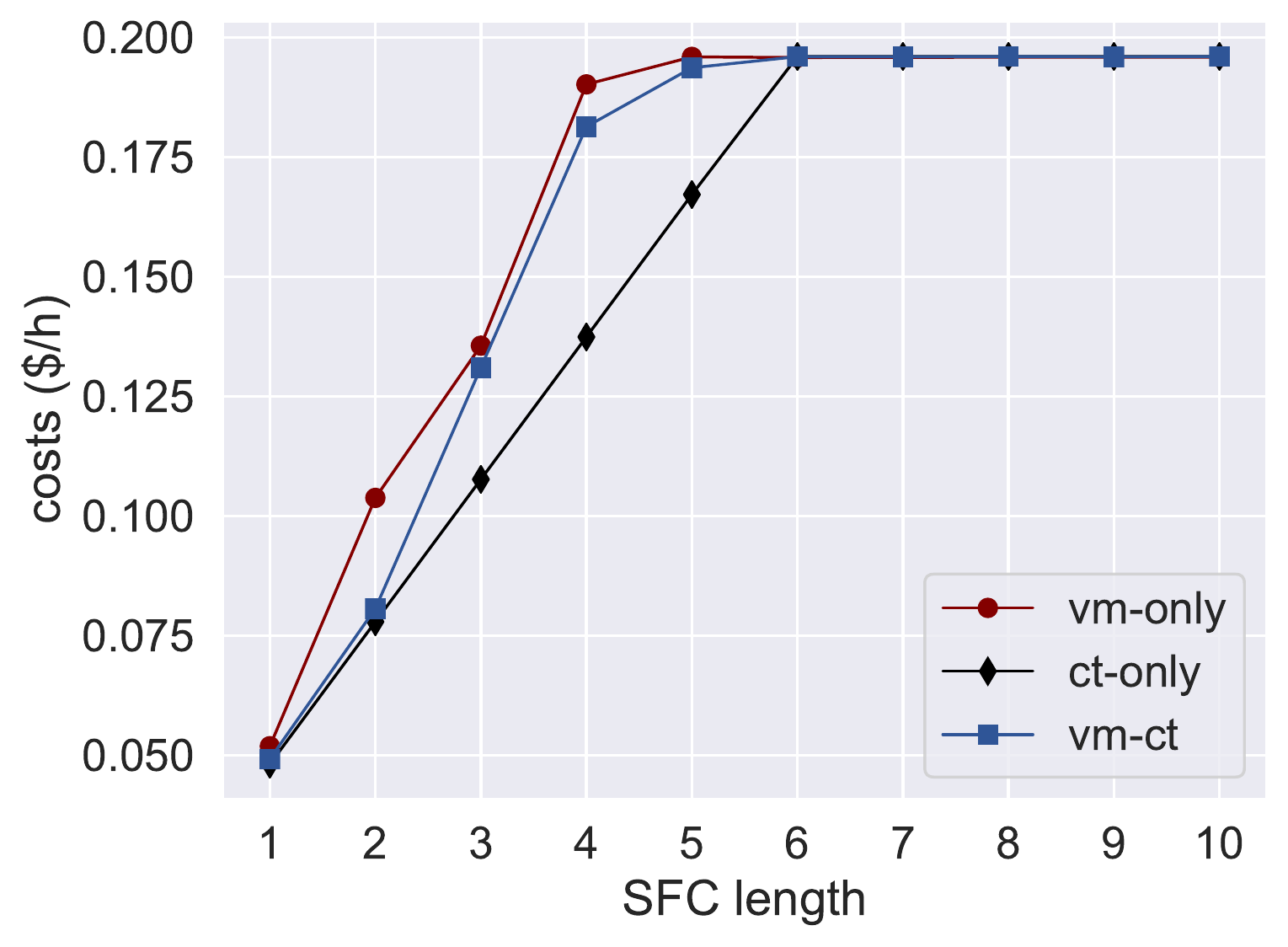}%
		\label{fig:7nodes_individual-costs_ox}}
	\hfil
	\subfloat[Cloud charges]{\includegraphics[width=0.50\columnwidth]{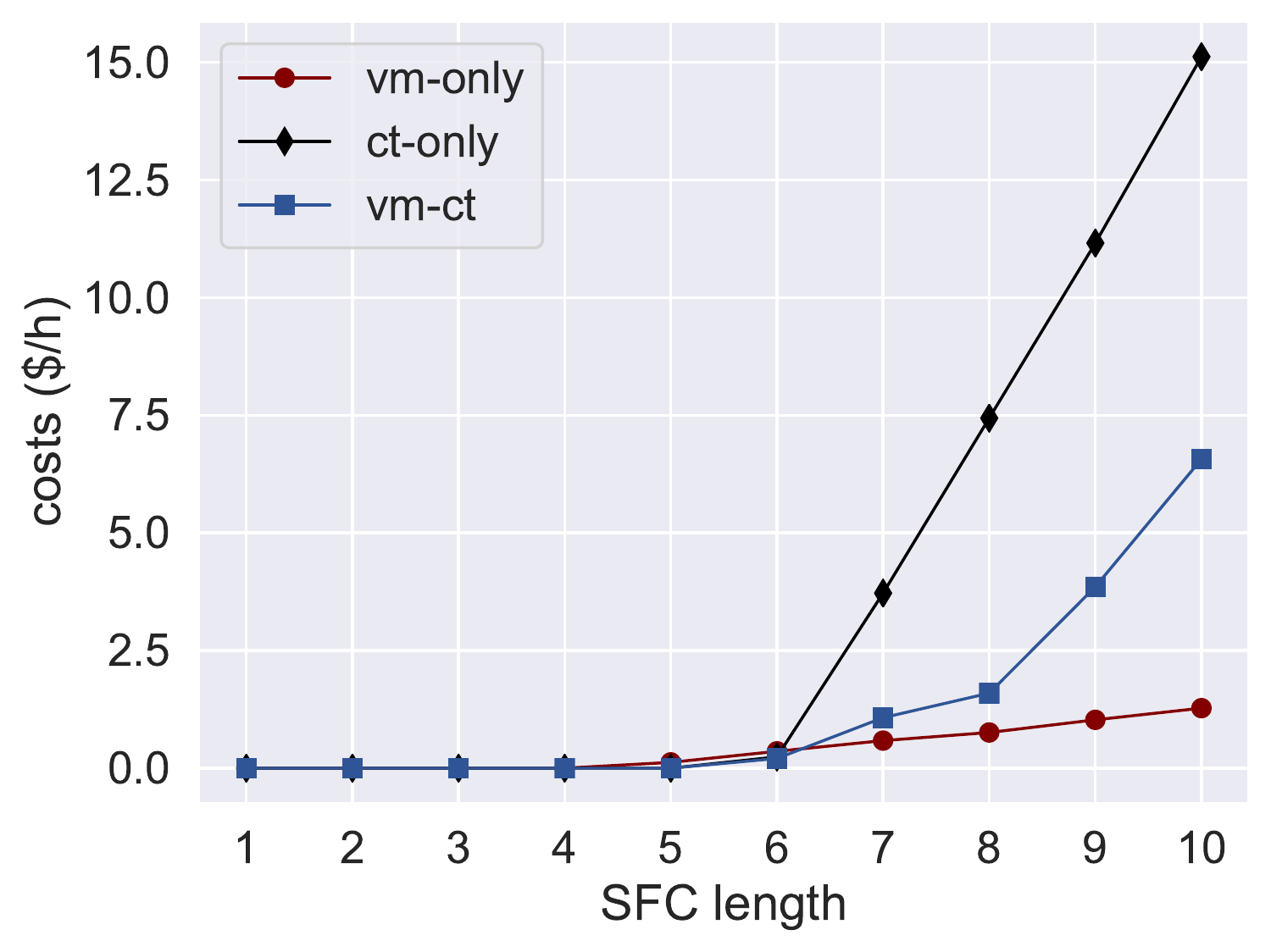}%
		\label{fig:7nodes_individual-costs_osv}}
	\hfil
	\subfloat[Penalty costs]{\includegraphics[width=0.50\columnwidth]{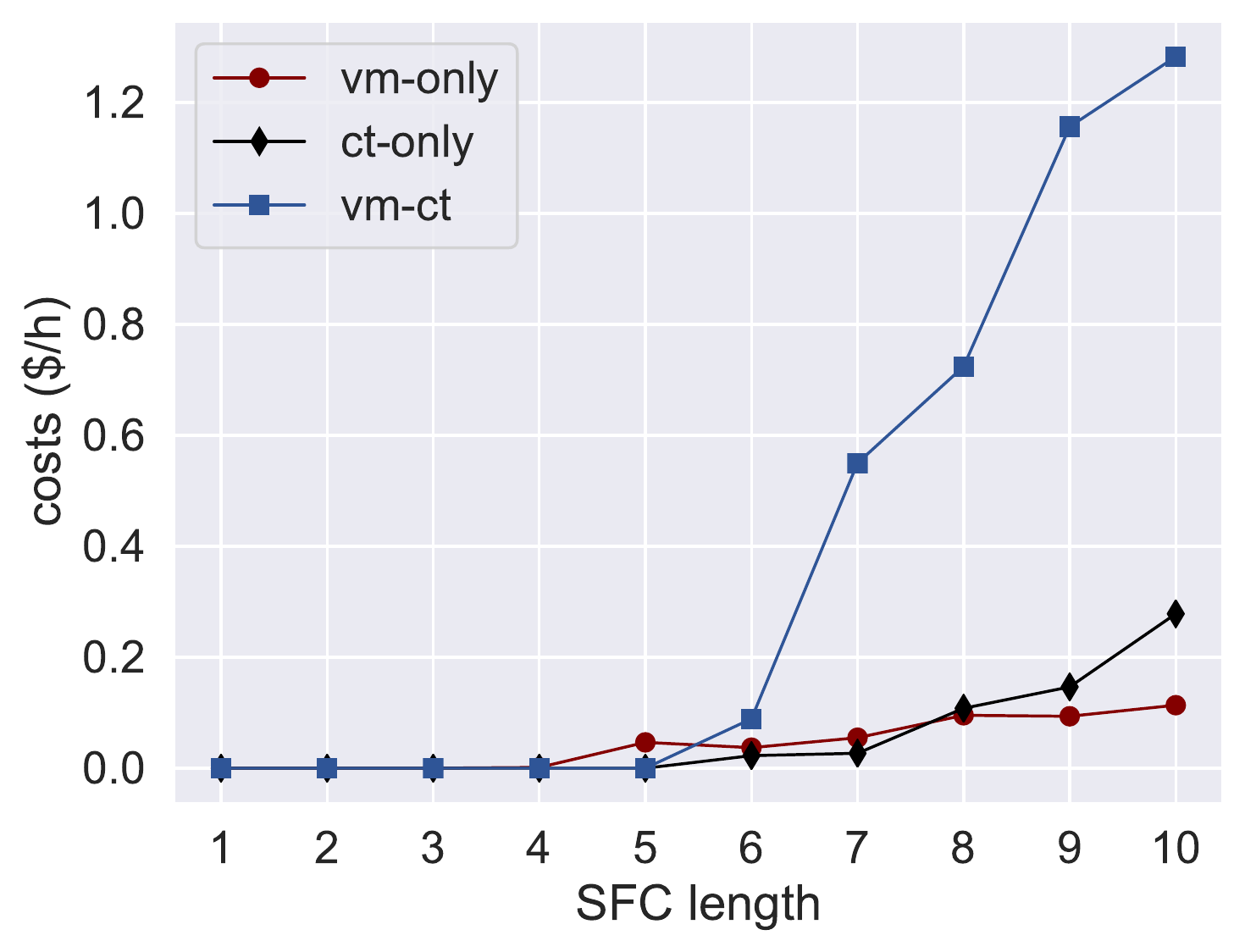}%
		\label{fig:7nodes_individual-costs_qsdp}}
	\caption{Comparison of total and individual costs from MILP model for different SFC lengths in network A.}
	\label{fig:7nodes_costs}
\end{figure*}

For each source-destination pair of nodes, 3 paths are pre-computed that do not
traverse the cloud node and 1 additional path that does. The path computation is
carried out is this way to make sure the model has enough freedom to allocate
all SFCs at the edge and at least there is one admissible path per SFC to
allocate VNFs in the cloud. We assume that every source-destination pair of
nodes instantiates independent SFCs and randomly generates between 1 and 3
traffic flows, each one with a bandwidth between [1, 20] units. Considering
$\Lambda$ as the total set of traffic demands in the network, for the initial
placement we only consider a subset $\Lambda'$ generated by using a selection
probability of $R=0.3$. In order to make a fair comparison between VMs and
containers, we assume that the VNFs themselves have the same parameters
irrespectively off the VM or container configurations; the only difference is
the overhead introduced by VMs ($\overhead$) and the charges applied when
allocating VNFs in the cloud ($\functioncharges$). The overhead $\overhead$ is
calculated as ~10\% of the maximum processing capacity $\maxcapserver$ that a
VNF can have \cite{Reddy2014}. Since $\maxcapserver$ is maximum possible
processing capacity that a VNF can handle, it is calculated based on the worst
case scenario, which is as the maximum number of traffic demands multiplied by
the maximum possible bandwidth and by the traffic load ratio $\loadratio$. The
rest of the parameters can be found in Table \ref{tab:vnf-parameters}. Then, for
every s-d combination (except the cloud node), one SFC is created to provision a
service. The length of the SFCs varies from one to ten VNFs and we compare the
cases where all VNFs in the SFCs are deployed only over VMs (\emph{\vmonly}),
only over containers (\emph{\ctonly}), or hybrid, i.e., when both types are
combined (\emph{\hybrid}). For hybrid SFCs, the VNFs are randomly assigned
either VMs or CTs, following a uniform distribution considering all possible
combinations. Finally, the penalty parameter $\penaltyparam$ is computed as 10\%
of the selling price for an specific SFC (see \cite{Racheg2017} or
\cite{Comcast}), which depends on the VNFs allocated to provide that specific
service. So, to calculate the selling price for every SFC, we consider the same
selling prices as for the charges of the cloud provider, so we use
$\functioncharges$ values as well for every VNF deployed at the edge. In the
proposed networks, for all SFCs the maximum propagation delay considered is
$\Dnet = 5 ms$ considering round trip time from both edge networks to the cloud.
With this propagation delay, the maximum service downtime caused by migrations
is expected to be $\migrationdelay = 27.5 ms$ \cite{Taleb2019}.

\subsection{MILP model}

\subsubsection{Optimization Costs}

Fig. \ref{fig:7nodes_costs-sfc} shows the total monetary cost (equation
\eqref{total-costs}) comparison for different SFC lengths in the small network
(Fig. \ref{fig:small-network}) using the MILP model. As it can be seen, until
SFC length 4, all three cases result in a comparable cost value, while between
length 4 and 6, \emph{\vmonly} gets higher costs than the other two cases. This
later case is interesting to show how the overhead introduced by VMs overload
the edge servers earlier than the other two cases, incurring into higher costs.
With length longer than 6, both \emph{\ctonly} and \emph{\hybrid} get much
higher costs than \emph{\vmonly}, being \emph{\ctonly} the worst case. This
behavior can be explained by the fact that above length 6, the edge network is
overloaded and the cloud needs to be used incurring into more charges due to
container charges being higher than VMs. It is remarkable than the hybrid case
gets relatively close values to the \emph{\ctonly} case. Looking into the costs
individually, Fig. \ref{fig:7nodes_individual-costs_ox}, Fig.
\ref{fig:7nodes_individual-costs_osv} and Fig.
\ref{fig:7nodes_individual-costs_qsdp} show the OPEX costs, cloud charges and
penalty costs. From the OPEX costs we confirm that in all cases the edge servers
are full with SFC length 6 and above. Here we notice that while \emph{\vmonly}
is the one incurring into more costs, \emph{\hybrid} costs are not much lower
and \emph{\ctonly} case is the one with lower costs until length 6. The opposite
happens with cloud charges where \emph{\ctonly} is the worse case by far
compared to the other two. If we take a look into the penalty costs results, we
can see how the hybrid \emph{\hybrid} has much higher above length 6 than the
other two cases. This latter case is interesting because we observe how the
model penalizes these costs which are compensated by highly reducing the cloud
charges which the respective values have higher impact on the total costs. This
effect on the penalty costs for the \emph{\hybrid} case is correlated with the
number of migrations (as shown later), which exceed the maximum allowed service
delay.

\begin{figure}[!t]
	\centering
	\subfloat[Migrations]{\includegraphics[width=0.50\columnwidth]{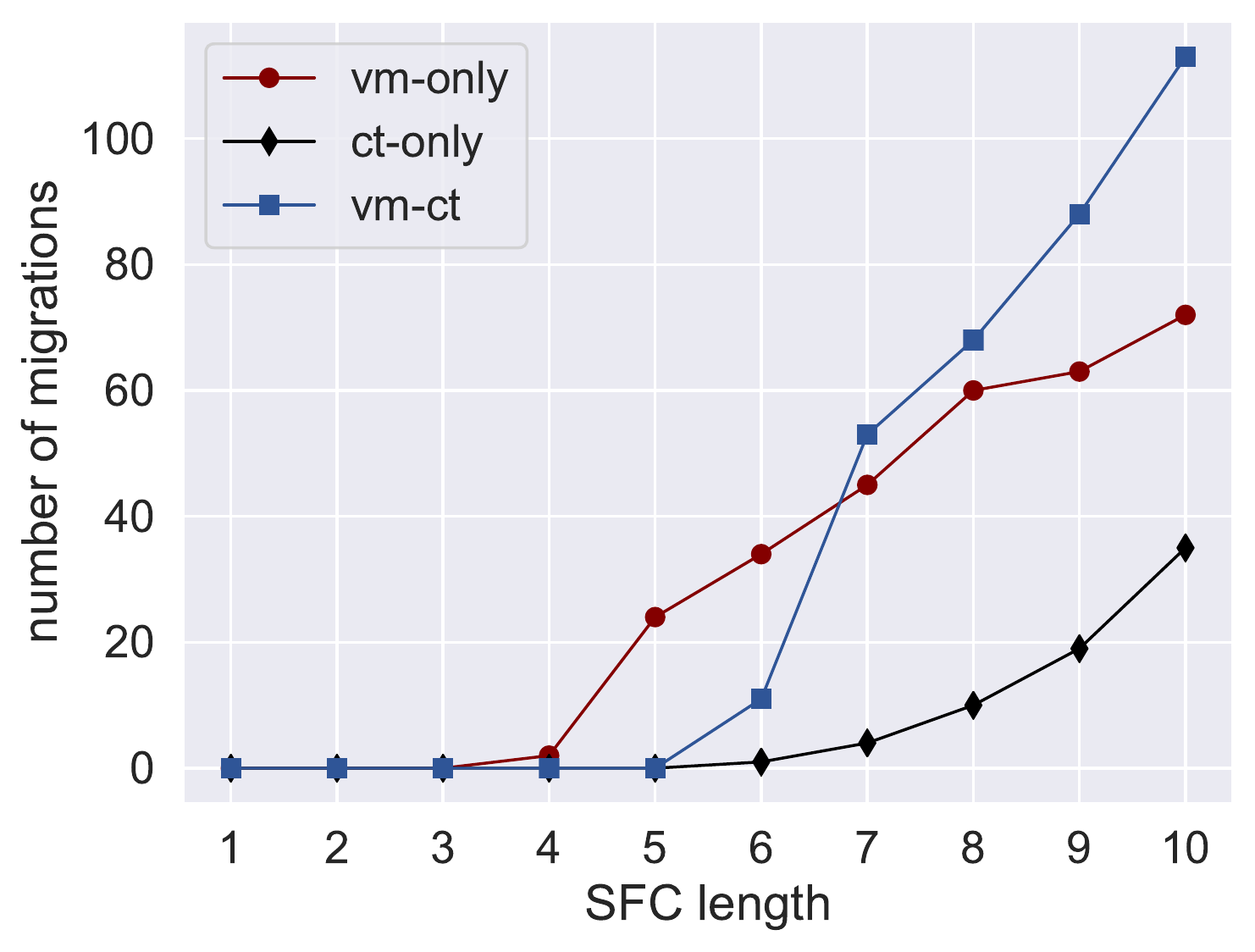}%
		\label{fig:7nodes_migrations}}
	\hfil
	\subfloat[Replications]{\includegraphics[width=0.50\columnwidth]{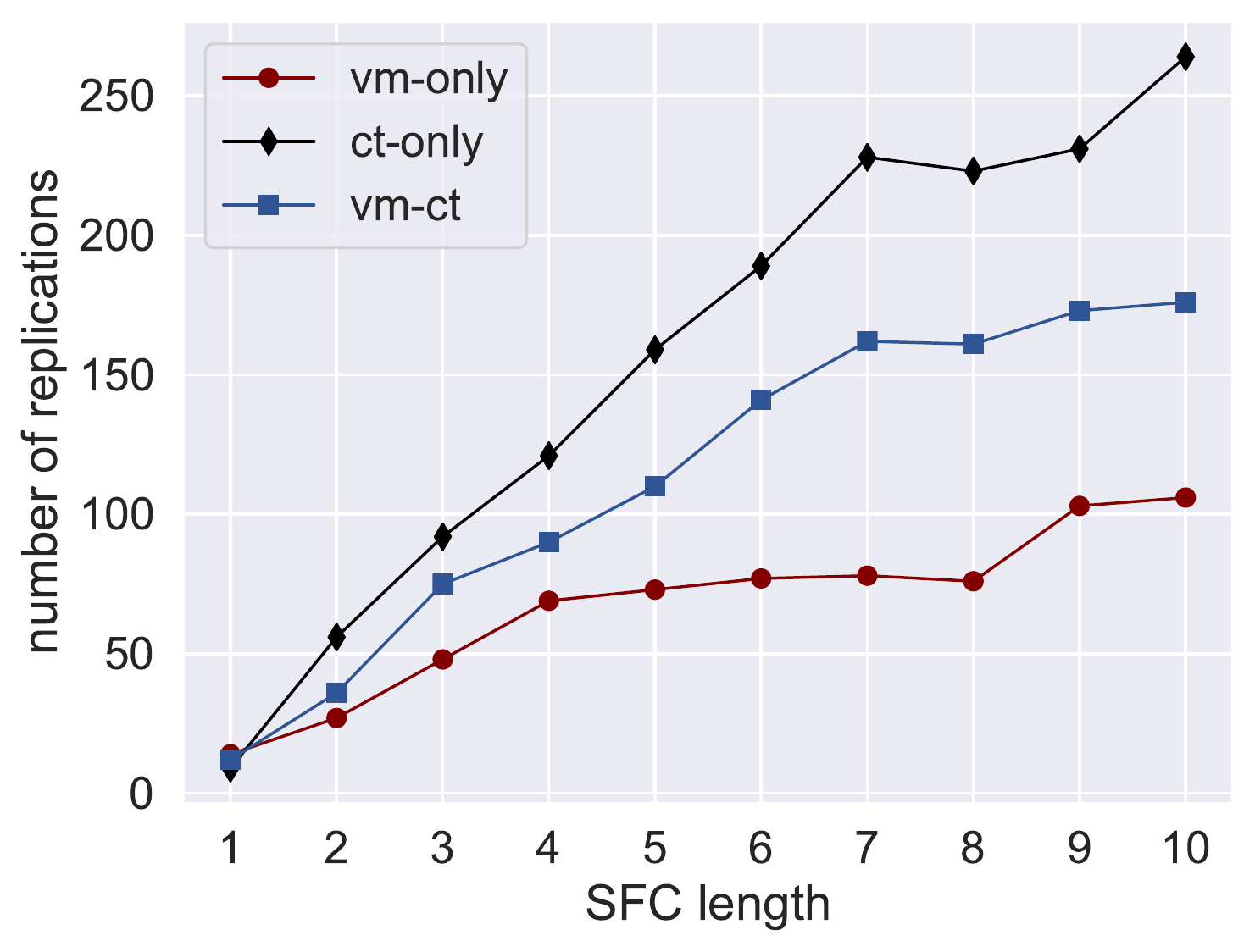}%
		\label{fig:7nodes_replications}}
	\caption{Total number of migrations and replications for different SFC lengths}
	\label{fig:7nodes_mgr_rep}
\end{figure}

\subsubsection{Migrations and Replications}

In order to better understand the costs results, we show in Fig.
\ref{fig:7nodes_migrations} and Fig. \ref{fig:7nodes_replications} the number of
VNF migrations and replications, respectively, which have a direct impact on the
costs. We can see that the model starts earlier to perform migrations in the
\emph{\vmonly} case, but with length 7 and above, the \emph{\hybrid} case
performs more migrations than any other case. While this behavior is
interesting, it is hard to explain why this occurs. One possible explanation
could be that these migrations occur within the edge network having a minimal
impact on the penalty costs, as we can see in the previous Fig.
\ref{fig:7nodes_individual-costs_qsdp}. On the other hand, for the number of
replications, the \emph{\vmonly} performs less replications than the other two
cases, being the \emph{\hybrid} case an intermediate solution. This result is as
expected since VMs introduce an overhead, so the model tries to minimize them.

\subsubsection{Resource Utilization and Service Delay}

\begin{figure}[!t]
	\centering
	\subfloat[Average link utilization]{\includegraphics[width=0.50\columnwidth]{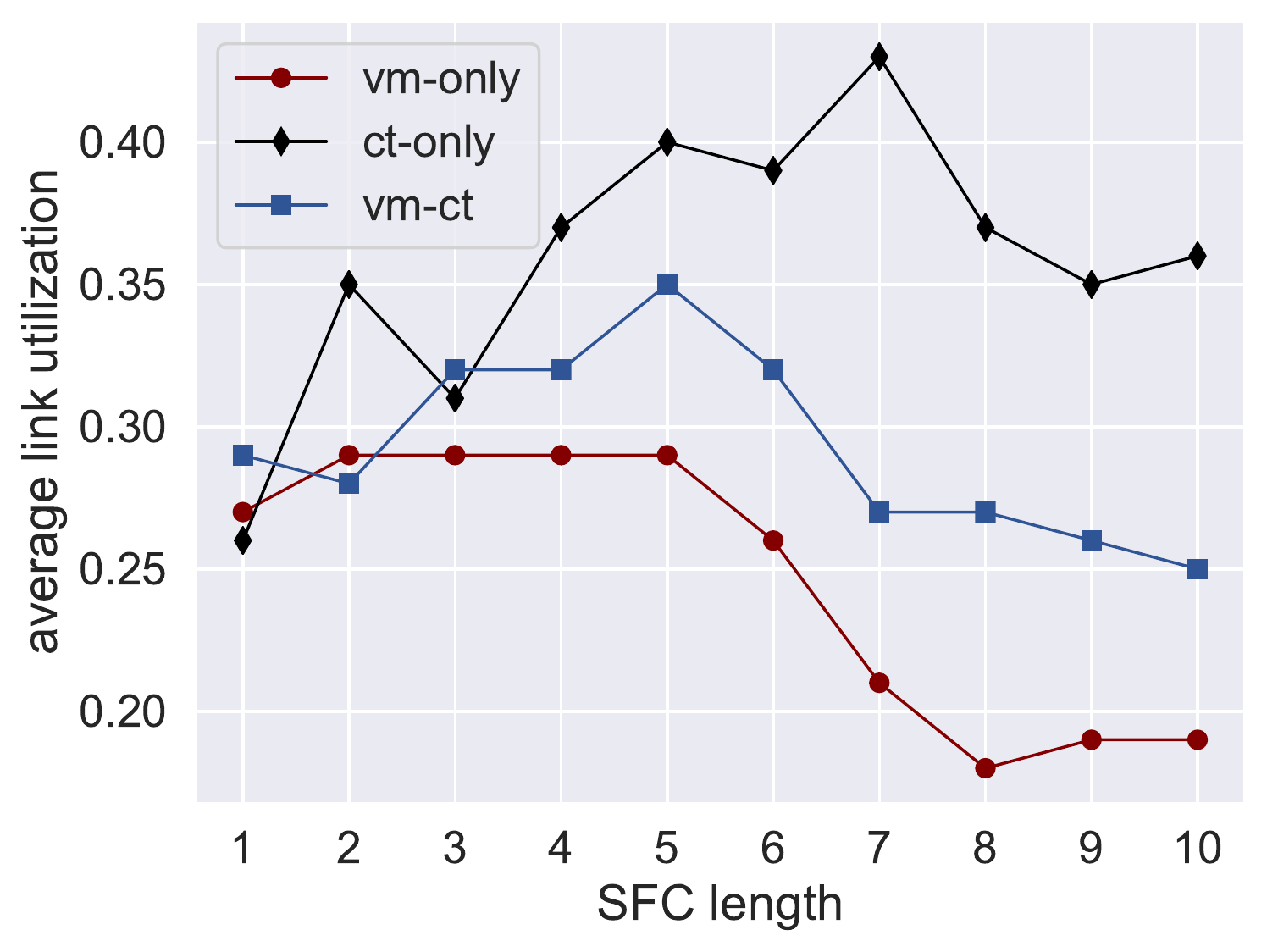}%
		\label{fig:7nodes_avg-lu}}
	\hfil
	\subfloat[Average server utilization]{\includegraphics[width=0.50\columnwidth]{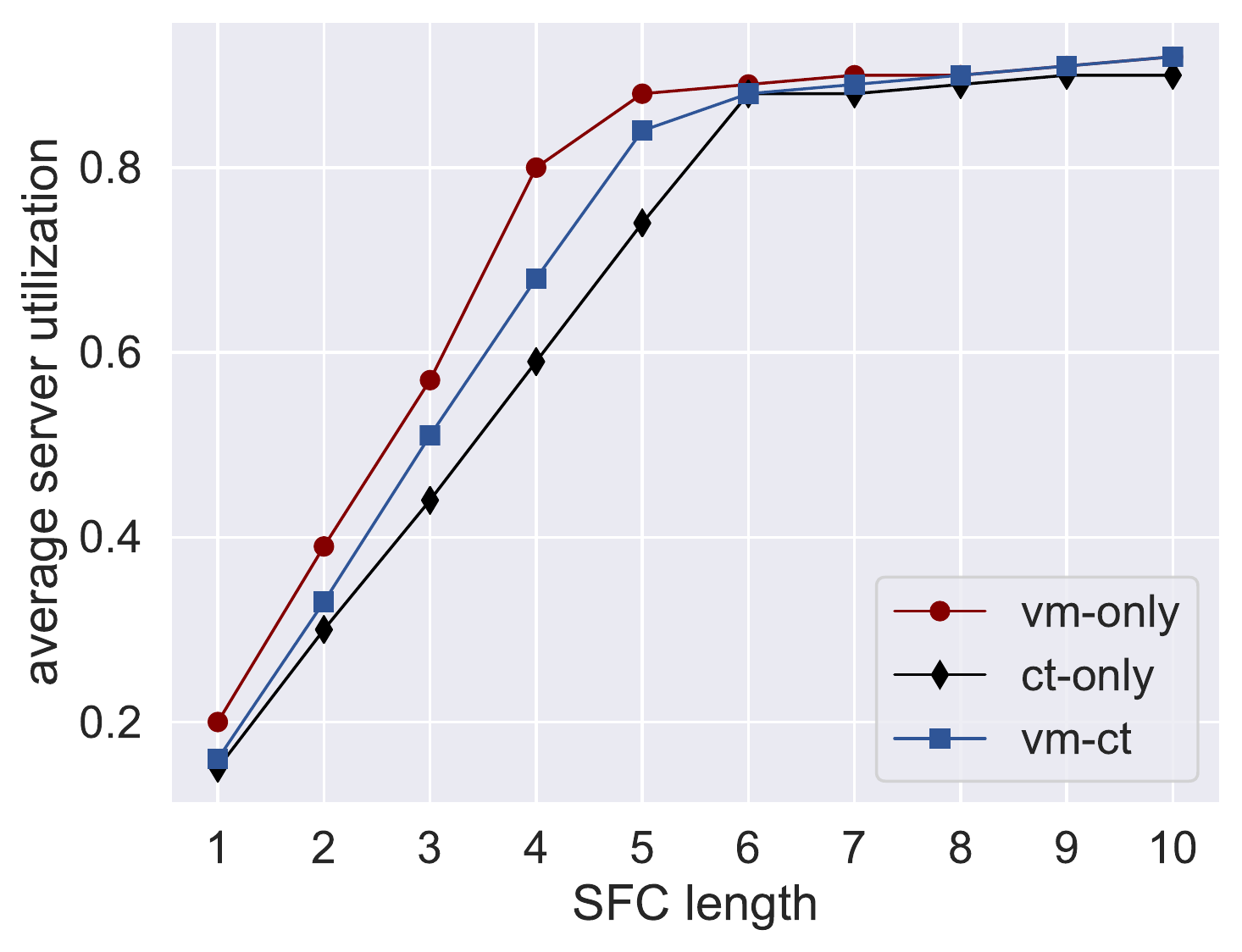}%
		\label{fig:7nodes_avg-xu}}
	\hfil
	\subfloat[Average service delay]{\includegraphics[width=0.50\columnwidth]{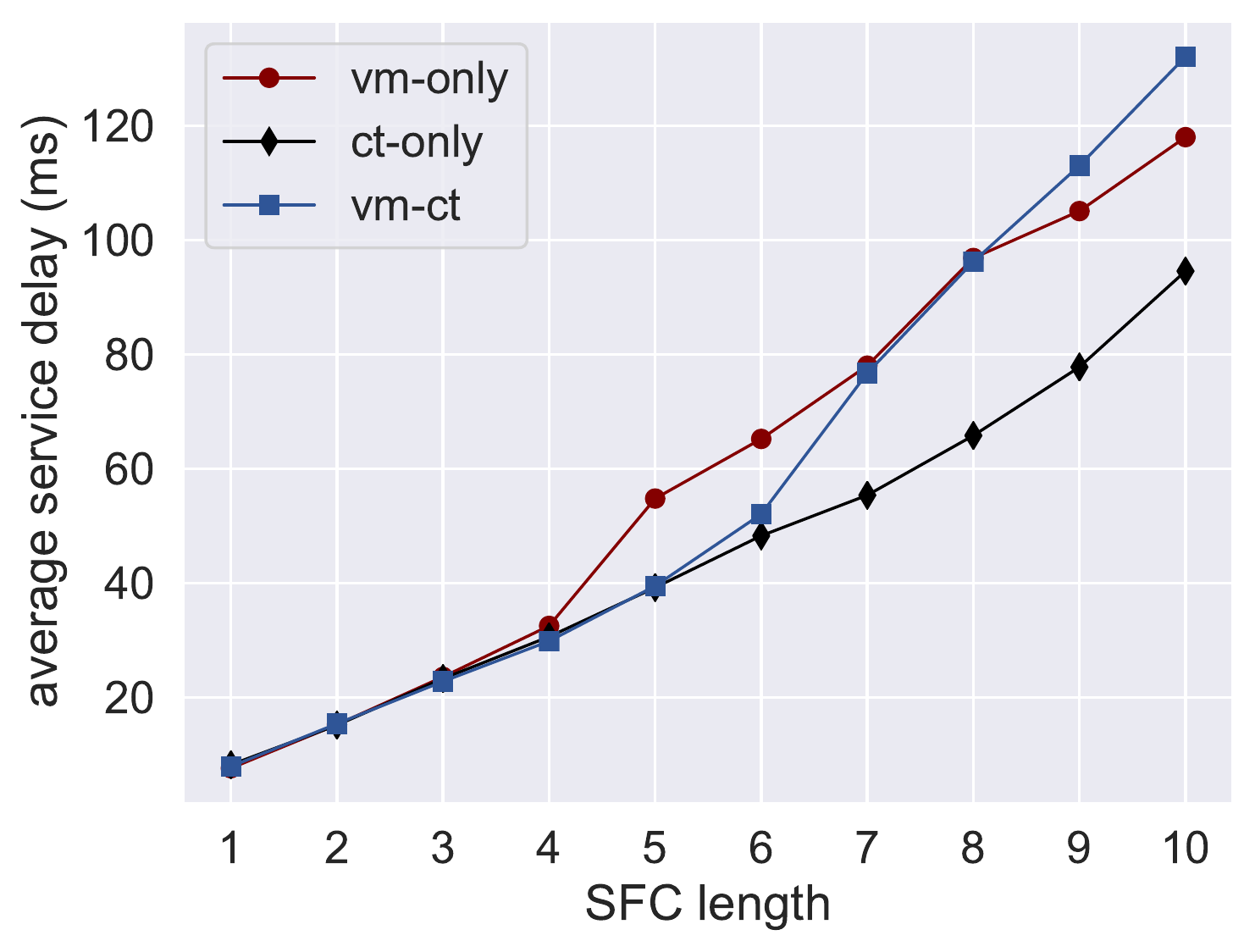}%
		\label{fig:7nodes_avg-sd}}
	\caption{Resource utilization and service delays for different SFC lengths}
	\label{fig:7nodes_resources}
\end{figure}

Fig. \ref{fig:7nodes_avg-lu} shows the average link utilization results for all
links in the network except those connecting to the cloud. Here, we can see how
the average network utilization for the \emph{\vmonly} is mostly lower than in
the other two cases, being \emph{\ctonly} the one with higher average and
\emph{\hybrid} case an intermediate solution. This is due to the synchronization
traffic between replicas that has to be added to the network load, and these
results are consistent with the number of replicas shown previously in Fig.
\ref{fig:7nodes_replications}. In all three cases, we can also see how above
length 5, the average decreases due to the migration of VNFs to the cloud, which
minimizes the traffic in the edge network. On the other side, the average server
utilization, shown in Fig. \ref{fig:7nodes_avg-xu}, confirms the OPEX costs
results, showing how the edge network is overloaded when the length is above 6.
It should be noted that even though the number of replicas is higher for
\emph{\ctonly}, this does not affect to the average server utilization since
there is no overhead introduced.

Fig. \ref{fig:7nodes_avg-sd} shows the average end-to-end service delay between
all SFCs deployed in the network. Here, \emph{\vmonly}  increases linearly with
the SFC length and is the one with comparably lower delay. While  \emph{\hybrid}
has a similar average delay, above length 6, the value increases faster, being
higher than \emph{\ctonly} case above length 9. This behavior is correlated to
the number of migrations that can occur, due to the impact of service
interruptions on the overall delay. While this tendency can be a concern,
considering that in this case the edge network is completely overloaded and,
therefore, requires heavy offloading to the cloud, the delays due to migrations
are not a relevant in the cloud environments.

\begin{figure}[!t]
	\centering
	\subfloat[Total costs with \vmonly]{\includegraphics[width=0.50\columnwidth]{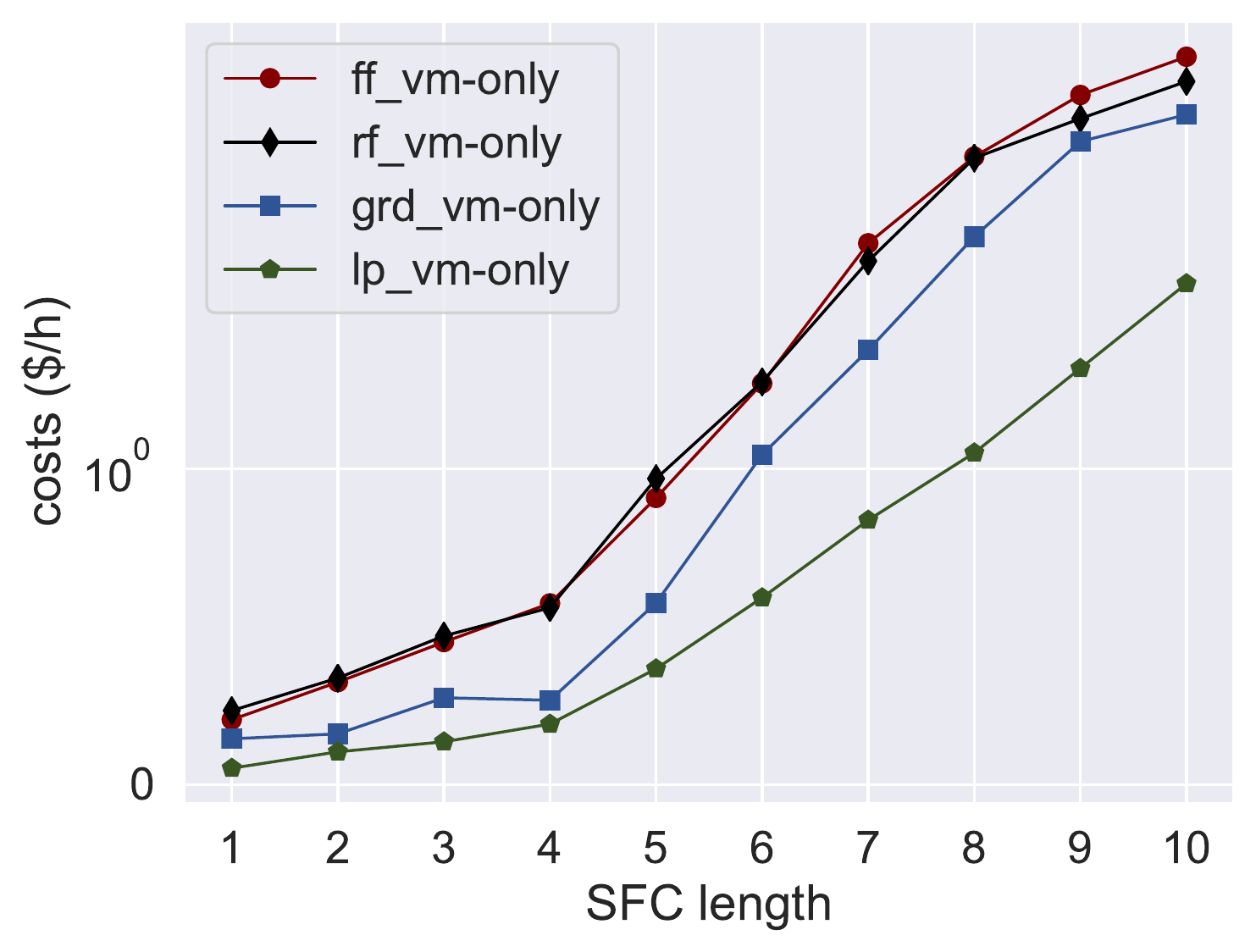}%
		\label{fig:7nodes_costs-alg_vm-only}}
	\hfil
	\subfloat[Total costs with \ctonly]{\includegraphics[width=0.50\columnwidth]{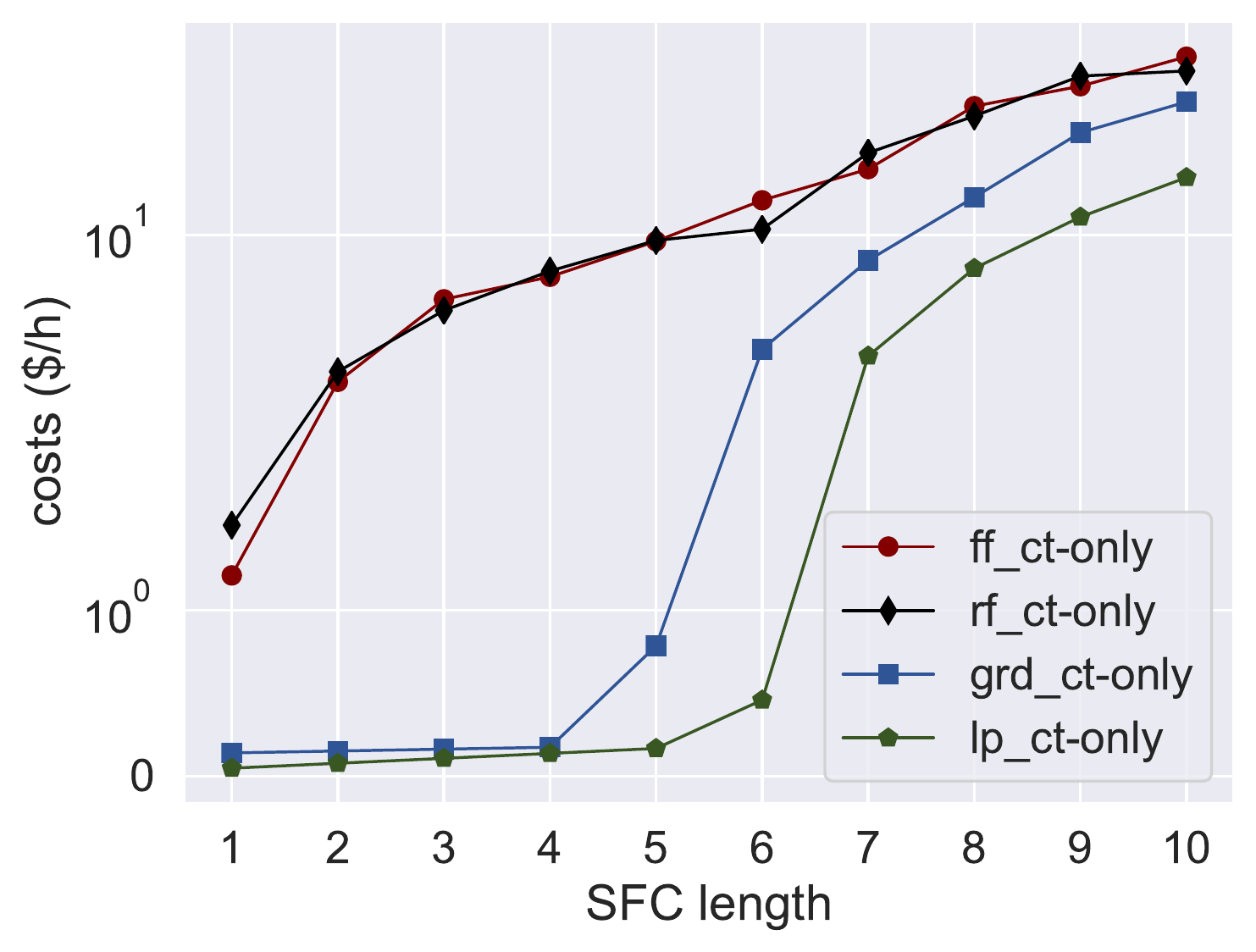}%
		\label{fig:7nodes_costs-alg_ct-only}}
	\hfil
	\subfloat[Total costs with \hybrid]{\includegraphics[width=0.50\columnwidth]{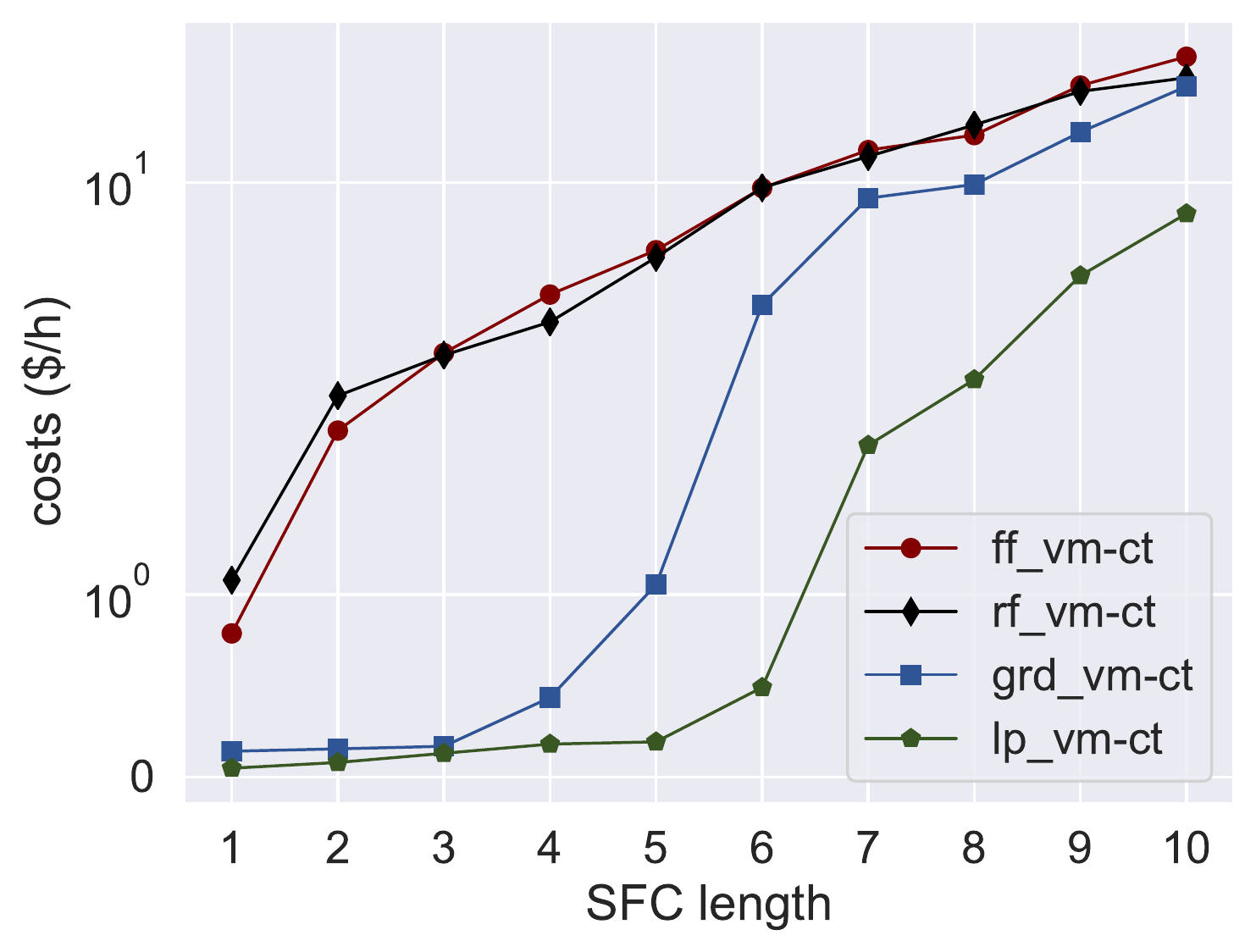}%
		\label{fig:7nodes_costs-alg_vm-ct}}
	\caption{Comparison of algorithms for different SFC lengths in small scale network.}
	\label{fig:7nodes_costs-alg}
\end{figure}

\begin{figure}[!t]
	\centering
	\subfloat[Total costs with vm-only]{\includegraphics[width=0.50\columnwidth]{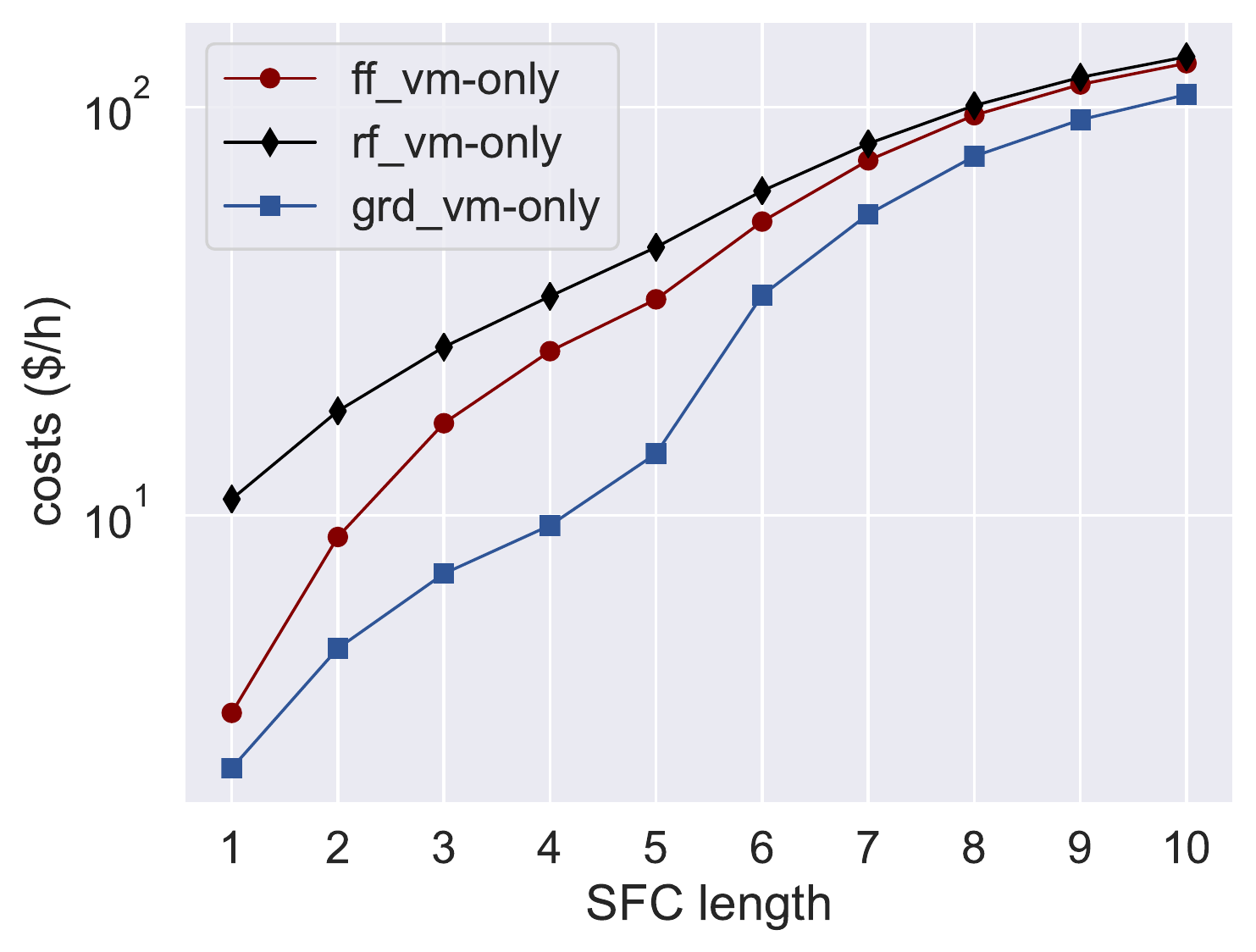}%
		\label{fig:palmetto_costs-alg_vm-only}}
	\hfil
	\subfloat[Total costs with \ctonly]{\includegraphics[width=0.50\columnwidth]{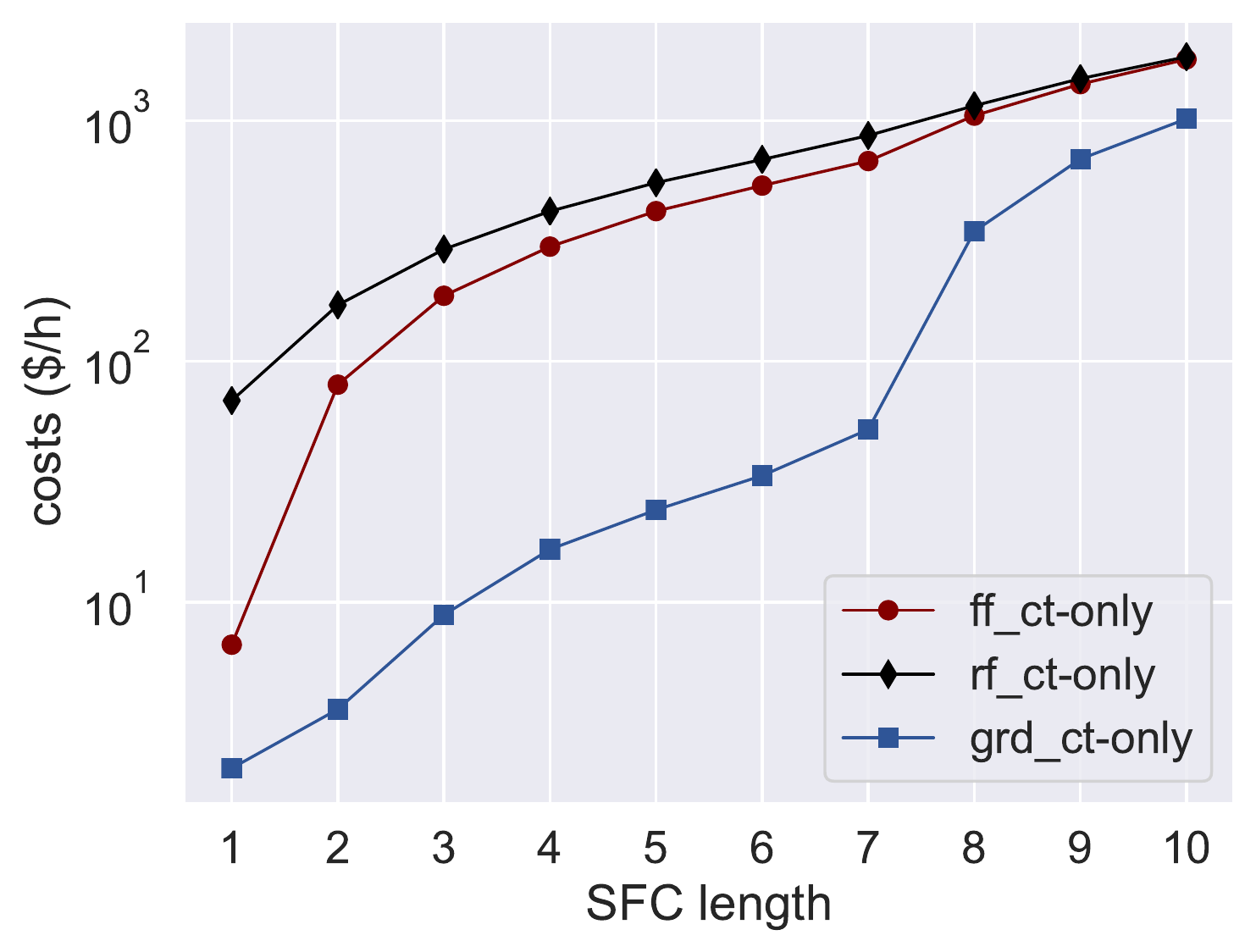}%
		\label{fig:palmetto_costs-alg_ct-only}}
	\hfil
	\subfloat[Total costs with \hybrid]{\includegraphics[width=0.50\columnwidth]{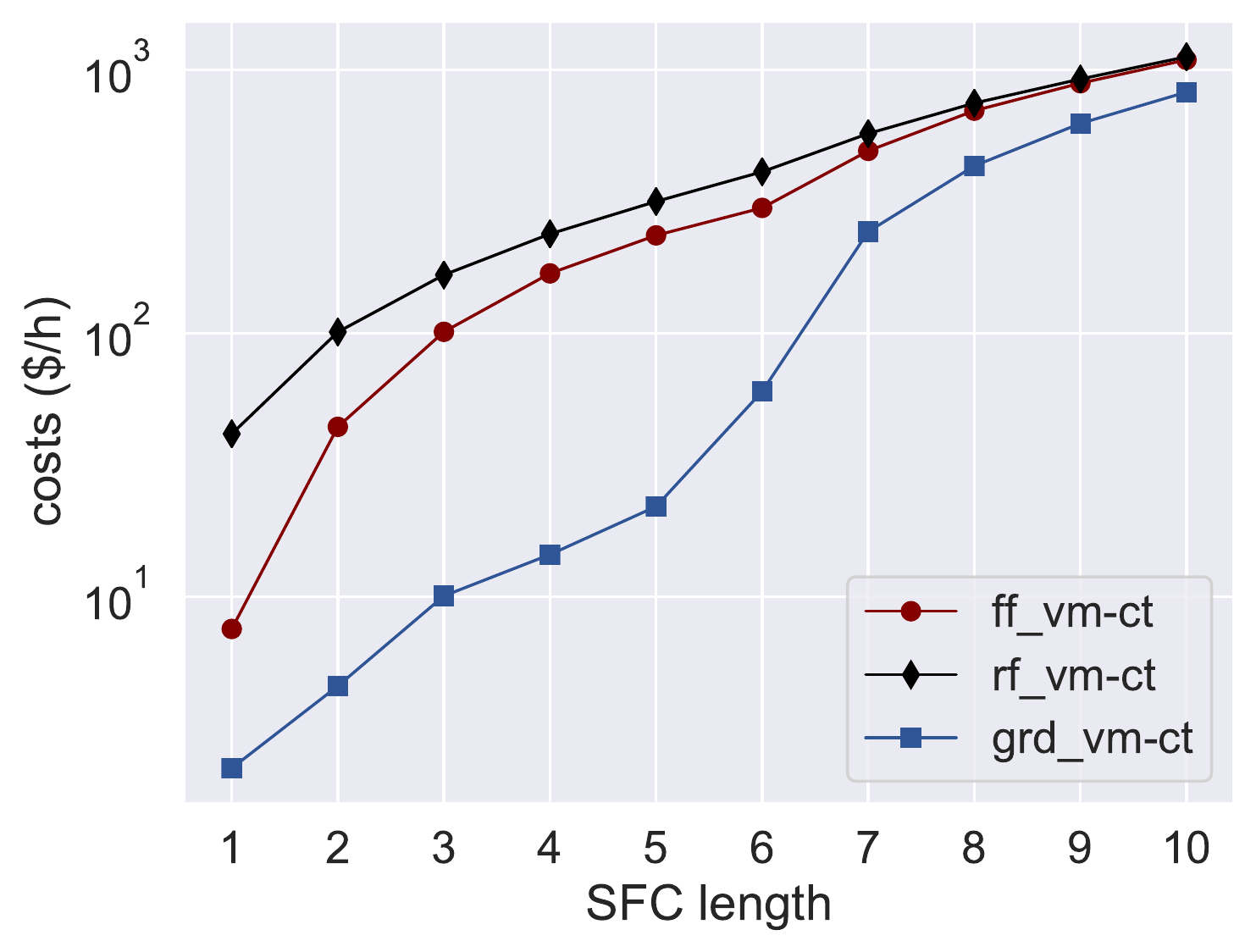}%
		\label{fig:palmetto_costs-alg_vm-ct}}
	\caption{Comparison of algorithms for different SFC lengths in large scale network.}
	\label{fig:palmetto_costs-alg}
\end{figure}

\subsection{Heuristics}

Here, we first compare the performance of the greedy algorithm compared to the
optimal solution and to both first-fit and random-fit for the small scale
network (Fig. \ref{fig:small-network}). Fig. \ref{fig:7nodes_costs-alg_vm-only},
Fig. \ref{fig:7nodes_costs-alg_ct-only} and Fig.
\ref{fig:7nodes_costs-alg_vm-ct} show the total monetary costs obtained using
first-fit (\emph{ff}), random-fit (\emph{rf}), greedy (\emph{grd}) and MILP
(\emph{lp}) for for the \emph{\vmonly}, \emph{\ctonly} and \emph{\hybrid} cases,
respectively. We can observe from all three cases that there is not much
difference between first-fit and random-fit algorithms. The fact that random-fit
behaves similar to first-fit can be due to the influence of the sequence order
of VNFs have in the network, which is relatively small. Even with random
placement, there is no much freedom of choice when placing VNFs, due to size.
Thus, the available servers are restricted by the placement of the previous and
next VNF of the chain in a specific path. In both cases, the fact that these
algorithms do not take into account where the VNFs were placed during the
initial placement, results in too many migrations and replications that increase
the penalty costs and resource utilization, respectively. On the other hand, we
can see how the greedy algorithm performs really close to the optimal solution
when the SFC length is up to 4, when the servers at the edge are still not
overloaded. For longer chains, the algorithm performs slightly worse due to the
edge servers are overloaded and the cloud must be used, leading to a situation
where the algorithm has less freedom for placements. In Fig.
\ref{fig:palmetto_costs-alg}, we show the same costs for the large network (see
Fig. \ref{fig:large-network}), with heuristics only. As in the case of smaller
network, we can see how the greedy algorithm works better for  \emph{\ctonly}
than for the \emph{\vmonly} case.

\subsection{Discussion and remarks}

From the cost optimization point of view, we can see how the placement of VNFs
on servers at the edge incurs lower costs when the ISP owns the server
infrastructure as compared to the case of using a third party cloud provider.
Specifically, the use of \ctvnf~reduces the total costs even in hybrid scenarios
with \vmvnf. In the situations when the use of the cloud is necessary,
\ctvnf~can result in excessive costs as compared to VMs only, but the
combination of both in a hybrid SFC can alleviate these costs. We also showed
how the preceding placement of VNFs in the network impacts the resulting the
costs for the subsequent reallocations, since the number of migrations are going
to affect the penalty costs, while the number of replications impact the
resource usage. We also observe that when the edge network has enough server
resources for allocation, greedy algorithm that performs close to the optimal
solution (MILP). One relevant extension of the model would be to consider the
sharing of VNFs of the same type between different SFCs, so it can  be used for
studying reliability \cite{engelmann2018}. Finally, while our model used generic
VNF parameters, and the results may change with real-world parameters, the model
can still provide some important clues on how the migration from VM to
containers, and their hybrid combination with VMs can affect the cost models
from the ISP's point of view.

\section{Conclusions}

We studied a novel problem of optimal placement of hybrid SFCs, a combination of
virtual machines and containers, from an Internet Service Provider (ISP) point
of view, in a generic edge and cloud continuum. To this end, we proposed a
Mixed-Integer Linear Programming model as well as a heuristic solution to solve
the cost optimization problem that considers three objectives unique to the
specific VM and container deployment in a carrier network: operational costs for
maintaining servers in the edge, costs of placing VNFs in third-party cloud
providers and penalty costs applied when SLA agreements are violated in terms of
end-to-end delay. We also proposed 2-phases optimization process to analyze the
effect on performance as a result of replications and migrations of VNFs. For
larger networks, we developed a greedy algorithm that performs close to the
optimal solution when there are enough free resources in the edge network. The
results have shown that from the cost optimization point of view, the placement
of VNFs on servers at the edge incurs lower costs when the ISP owns the server
infrastructure compared to the case of using a third party cloud provider. As
future work, we plan to extend the model for supporting shared VNFs and evaluate
other kind scenarios where the objective if focused on the resource utilization.

\ifCLASSOPTIONcaptionsoff
	\newpage
\fi

\bibliographystyle{IEEEtran}
\bibliography{references}

\begin{thebibliography}{10}
\providecommand{\url}[1]{#1}
\csname url@samestyle\endcsname
\providecommand{\newblock}{\relax}
\providecommand{\bibinfo}[2]{#2}
\providecommand{\BIBentrySTDinterwordspacing}{\spaceskip=0pt\relax}
\providecommand{\BIBentryALTinterwordstretchfactor}{4}
\providecommand{\BIBentryALTinterwordspacing}{\spaceskip=\fontdimen2\font plus
\BIBentryALTinterwordstretchfactor\fontdimen3\font minus
  \fontdimen4\font\relax}
\providecommand{\BIBforeignlanguage}[2]{{%
\expandafter\ifx\csname l@#1\endcsname\relax
\typeout{** WARNING: IEEEtran.bst: No hyphenation pattern has been}%
\typeout{** loaded for the language `#1'. Using the pattern for}%
\typeout{** the default language instead.}%
\else
\language=\csname l@#1\endcsname
\fi
#2}}
\providecommand{\BIBdecl}{\relax}
\BIBdecl

\bibitem{ETSI2016}
ETSI, ``{Network Functions Virtualisation (NFV); Reliability; Report on Models
  and Features for End-to-End Reliability},'' 2016.

\bibitem{aws}
\BIBentryALTinterwordspacing
Amazon, ``{AWS Lambda}.'' [Online]. Available:
  \url{https://aws.amazon.com/lambda/}
\BIBentrySTDinterwordspacing

\bibitem{Alvarez2019b}
F.~Alvarez, D.~Breitgand, D.~Griffin, P.~Andriani, S.~Rizou, N.~Zioulis,
  F.~Moscatelli, J.~Serrano, M.~Keltsch, P.~Trakadas, T.~K. Phan, A.~Weit,
  U.~Acar, O.~Prieto, F.~Iadanza, G.~Carrozzo, H.~Koumaras, D.~Zarpalas, and
  D.~Jimenez, ``{An Edge-to-Cloud Virtualized Multimedia Service Platform for
  5G Networks},'' \emph{IEEE Transactions on Broadcasting}, vol.~65, no.~2, pp.
  369--380, 2019.

\bibitem{Tajiki2017}
\BIBentryALTinterwordspacing
M.~M. Tajiki, S.~Salsano, L.~Chiaraviglio, M.~Shojafar, and B.~Akbari, ``{Joint
  Energy Efficient and QoS-aware Path Allocation and VNF Placement for Service
  Function Chaining},'' \emph{IEEE Transactions on Network and Service
  Management}, vol.~PP, no.~c, p.~1, 2017. [Online]. Available:
  \url{http://arxiv.org/abs/1710.02611}
\BIBentrySTDinterwordspacing

\bibitem{Wang2016a}
L.~Wang, Z.~Lu, X.~Wen, R.~Knopp, and R.~Gupta, ``{Joint Optimization of
  Service Function Chaining and Resource Allocation in Network Function
  Virtualization},'' \emph{IEEE Access}, vol.~4, pp. 8084--8094, 2016.

\bibitem{Basta2017}
A.~Basta, A.~Blenk, K.~Hoffmann, H.~J. Morper, M.~Hoffmann, and W.~Kellerer,
  ``{Towards a Cost Optimal Design for a 5G Mobile Core Network based on SDN
  and NFV},'' \emph{IEEE Transactions on Network and Service Management}, vol.
  4537, no.~c, pp. 1--14, 2017.

\bibitem{Laghrissi2019}
A.~Laghrissi and T.~Taleb, ``{A Survey on the Placement of Virtual Resources
  and Virtual Network Functions},'' \emph{IEEE Communications Surveys and
  Tutorials}, vol.~21, no.~2, pp. 1409--1434, 2019.

\bibitem{Xia2016}
J.~Xia, D.~Pang, Z.~Cai, M.~Xu, and G.~Hu, ``{Reasonably Migrating Virtual
  Machine in NFV-Featured Networks},'' \emph{IEEE International Conference on
  Computer and Information Technology (CIT)}, 2016.

\bibitem{Xia2016a}
J.~Xia, Z.~Cai, and M.~Xu, ``{Optimized Virtual Network Functions Migration for
  NFV},'' \emph{IEEE 22nd International Conference on Parallel and Distributed
  Systems Optimized}, 2016.

\bibitem{Gember-Jacobson2014}
\BIBentryALTinterwordspacing
A.~Gember-Jacobson, R.~Viswanathan, C.~Prakash, R.~Grandl, J.~Khalid, S.~Das,
  and A.~Akella, ``{OpenNF: Enabling Innovation in Network Function Control},''
  \emph{SIGCOMM Comput. Commun. Rev.}, vol.~44, no.~4, pp. 163--174, 2014.
  [Online]. Available: \url{http://doi.acm.org/10.1145/2740070.2626313}
\BIBentrySTDinterwordspacing

\bibitem{Ghaznavi2015}
M.~Ghaznavi, A.~Khan, N.~Shahriar, K.~Alsubhi, R.~Ahmed, and R.~Boutaba,
  ``{Elastic virtual network function placement},'' \emph{IEEE 4th
  International Conference on Cloud Networking (CloudNet)}, 2015.

\bibitem{Hawilo2017}
H.~Hawilo, M.~Jammal, and A.~Shami, ``{Orchestrating network function
  virtualization platform: Migration or re-instantiation?}'' \emph{Proceedings
  of the 2017 IEEE 6th International Conference on Cloud Networking, CloudNet
  2017}, 2017.

\bibitem{Zou2018}
J.~Zou, W.~Li, J.~Wang, Q.~Qi, and H.~Sun, ``{NFV Orchestration and Rapid
  Migration Based on Elastic Virtual Network and Container Technology},''
  \emph{2018 IEEE International Conference on Information Communication and
  Signal Processing, ICICSP 2018}, no. Icsp, pp. 6--10, 2018.

\bibitem{Eramo2017}
V.~Eramo, E.~Miucci, M.~Ammar, and F.~G. Lavacca, ``{An Approach for Service
  Function Chain Routing and Virtual Function Network Instance Migration in
  Network Function Virtualization Architectures},'' \emph{IEEE/ACM Transactions
  on Networking}, 2017.

\bibitem{michael2016}
\BIBentryALTinterwordspacing
K.~S. {Michael Till Beck, Juan Felipe Botero}, M.~T. Beck, J.~F. Botero, and
  K.~S. {Michael Till Beck, Juan Felipe Botero}, ``{Resilient Allocation of
  Service Function Chains},'' in \emph{IEEE Conference on Network Function
  Virtualization and Software Defined Networks (NFV-SDN)}, 2016. [Online].
  Available: \url{http://ijsrm.in/index.php/ijsrm/article/view/1250}
\BIBentrySTDinterwordspacing

\bibitem{Ding2017}
W.~Ding, H.~Yu, and S.~Luo, ``{Enhancing the reliability of services in NFV
  with the cost-efficient redundancy scheme},'' \emph{IEEE International
  Conference on Communications}, vol.~1, 2017.

\bibitem{Qu2017}
L.~Qu, C.~Assi, K.~Shaban, and M.~Khabbaz, ``{A Reliability-Aware Network
  Service Chain Provisioning with Delay Guarantees in NFV-Enabled Enterprise
  Datacenter Networks},'' 2017.

\bibitem{Alleg2018a}
A.~Alleg, T.~Ahmed, M.~Mosbah, R.~Riggio, and R.~Boutaba, ``{Delay-aware VNF
  placement and chaining based on a flexible resource allocation approach},''
  \emph{2017 13th International Conference on Network and Service Management,
  CNSM 2017}, vol. 2018-Janua, pp. 1--7, 2018.

\bibitem{Carpio2017a}
F.~Carpio, S.~Dhahri, and A.~Jukan, ``{VNF placement with replication for Load
  balancing in NFV networks},'' in \emph{IEEE International Conference on
  Communications}, 2017.

\bibitem{Carpio2017b}
F.~Carpio, W.~Bziuk, and A.~Jukan, ``{Replication of Virtual Network Functions:
  Optimizing link utilization and resource costs},'' \emph{2017 40th
  International Convention on Information and Communication Technology,
  Electronics and Microelectronics, MIPRO 2017 - Proceedings}, pp. 521--526,
  2017.

\bibitem{Huang2018}
M.~Huang, W.~Liang, Y.~Ma, and S.~Guo, ``{Throughput maximization of
  delay-sensitive request admissions via virtualized network function
  placements and migrations},'' \emph{IEEE International Conference on
  Communications}, vol. 2018-May, no.~c, 2018.

\bibitem{Carpio2018}
F.~Carpio, A.~Jukan, and R.~Pries, ``{Balancing the Migration of Virtual
  Network Functions with Replications in Data Centers},'' \emph{NOMS 2018 -
  2018 IEEE/IFIP Network Operations and Management Symposium}, pp. 1--8, 2018.

\bibitem{Filipe2019}
J.~B. Filipe, F.~Meneses, A.~U. Rehman, D.~Corujo, and R.~L. Aguiar, ``{A
  Performance Comparison of Containers and Unikernels for Reliable 5G
  Environments},'' \emph{2019 15th International Conference on the Design of
  Reliable Communication Networks, DRCN 2019}, pp. 99--106, 2019.

\bibitem{Riggio2018}
R.~Riggio, S.~N. Khan, T.~Subramanya, I.~G.~B. Yahia, and D.~Lopez,
  ``{LightMANO: Converging NFV and SDN at the edges of the network},''
  \emph{IEEE/IFIP Network Operations and Management Symposium: Cognitive
  Management in a Cyber World, NOMS 2018}, pp. 1--9, 2018.

\bibitem{Sewak2018}
\BIBentryALTinterwordspacing
M.~Sewak, ``{Winning in the Era of Serverless Computing and Function as a
  Service - IEEE Conference Publication},'' \emph{2018 3rd International
  Conference for Convergence in Technology (I2CT)}, pp. 1--5, 2018. [Online].
  Available: \url{https://ieeexplore.ieee.org/abstract/document/8529465}
\BIBentrySTDinterwordspacing

\bibitem{Sheoran2017}
A.~Sheoran, X.~Bu, L.~Cao, P.~Sharma, and S.~Fahmy, ``{An empirical case for
  container-driven fine-grained VNF resource flexing},'' \emph{2016 IEEE
  Conference on Network Function Virtualization and Software Defined Networks,
  NFV-SDN 2016}, pp. 121--127, 2017.

\bibitem{OSM2019}
O.~E. U.~A. Group, ``{OSM White Paper},'' 2019.

\bibitem{Cziva2018}
R.~Cziva, C.~Anagnostopoulos, and D.~P. Pezaros, ``{Dynamic, Latency-Optimal
  vNF Placement at the Network Edge},'' \emph{Proceedings - IEEE INFOCOM}, vol.
  2018-April, pp. 693--701, 2018.

\bibitem{Soni2017}
H.~Soni, W.~Dabbous, T.~Turletti, and H.~Asaeda, ``{NFV-based scalable
  guaranteed-bandwidth multicast service for software defined ISP networks},''
  \emph{IEEE Transactions on Network and Service Management}, vol.~14, no.~4,
  pp. 1157--1170, 2017.

\bibitem{Boubendir2017}
A.~Boubendir, E.~Bertin, and N.~Simoni, ``{On-demand, dynamic and at-the-edge
  VNF deployment model application to Web Real-Time Communications},''
  \emph{2016 12th International Conference on Network and Service Management,
  CNSM 2016 and Workshops, 3rd International Workshop on Management of SDN and
  NFV, ManSDN/NFV 2016, and International Workshop on Green ICT and Smart
  Networking, GISN 2016}, pp. 318--323, 2017.

\bibitem{Ma2018}
Y.~Ma, W.~Liang, M.~Huang, and S.~Guo, ``{Profit Maximization of NFV-Enabled
  Request Admissions in SDNs},'' \emph{2018 IEEE Global Communications
  Conference, GLOBECOM 2018 - Proceedings}, 2018.

\bibitem{Racheg2017}
W.~Racheg, N.~Ghrada, and M.~F. Zhani, ``{Profit-driven resource provisioning
  in NFV-based environments},'' \emph{IEEE International Conference on
  Communications}, pp. 1--7, 2017.

\bibitem{Bari2016}
M.~F. Bari, S.~R. Chowdhury, R.~Ahmed, R.~Boutaba, and O.~C. M.~B. Duarte,
  ``{Orchestrating Virtualized Network Functions},'' \emph{IEEE Transactions on
  Network and Service Management}, 2016.

\bibitem{Taleb2019}
T.~Taleb, A.~Ksentini, and P.~A. Frangoudis, ``{Follow-me cloud: When cloud
  services follow mobile users},'' \emph{IEEE Transactions on Cloud Computing},
  vol.~7, no.~2, pp. 369--382, 2019.

\bibitem{Bulut2015}
\BIBentryALTinterwordspacing
A.~Bulut and T.~Ralphs, ``{On the Complexity of Inverse Mixed Integer Linear
  Optimization},'' pp. 1--18, 2015. [Online]. Available:
  \url{http://coral.ie.lehigh.edu/{~}ted/files/papers/InverseMILP15.pdf}
\BIBentrySTDinterwordspacing

\bibitem{Psu2011}
D.~Psu, ``{ENERGY STAR {\textregistered} Power and Performance Data Sheet},''
  pp. 7--9, 2011.

\bibitem{Reddy2014}
P.~V.~V. Reddy and L.~Rajamani, ``{Virtualization overhead findings of four
  hypervisors in the CloudStack with SIGAR},'' \emph{2014 4th World Congress on
  Information and Communication Technologies, WICT 2014}, pp. 140--145, 2014.

\bibitem{Comcast}
{Comcast Technology Solutions}, ``{Service Level Agreement for Wholesale
  Dedicated Internet Last},'' 2009.

\bibitem{engelmann2018}
A.~{Engelmann} and A.~{Jukan}, ``A reliability study of parallelized vnf
  chaining,'' in \emph{2018 IEEE International Conference on Communications
  (ICC)}, 2018, pp. 1--6.

\end{thebibliography}

\end{document}